\documentclass[sn-mathphys-num]{sn-jnl}

\usepackage{float}
\usepackage{pgfplots}
\usepackage{graphicx}%
\usepackage{multirow}%
\usepackage{amsmath,amssymb,amsfonts}%
\usepackage{amsthm}%
\usepackage{mathrsfs}%
\usepackage[title]{appendix}%
\usepackage{xcolor}%
\usepackage{textcomp}%
\usepackage{manyfoot}%
\usepackage{booktabs}%
\usepackage{algorithm}%
\usepackage{algorithmicx}%
\usepackage{algpseudocode}%
\usepackage{listings}%
\usepackage{enumitem}

\theoremstyle{thmstyleone}%
\newtheorem{theorem}{Theorem}
\newtheorem{proposition}[theorem]{Proposition}%
\newtheorem{corollary}[theorem]{Corollary}%
\newtheorem{problem}{Problem}{}{}
\newtheorem{form}{Formulation}{}{}

\theoremstyle{thmstyletwo}%
\newtheorem{example}{Example}%

\theoremstyle{thmstylethree}%
\newtheorem{definition}{Definition}%

\begin{document}

\title[On properties driving diversity index selection]{On properties driving diversity index selection}


\author*[1]{\fnm{Martin} \sur{Frohn}}\email{martin.frohn@maastrichtuniversity.nl}

\author[2]{\fnm{Kerry} \sur{Manson}}\email{kerrymanson320@gmail.com}

\affil*[1]{\orgdiv{Department of Advanced Computing Sciences}, \orgname{Maastricht University}, \orgaddress{\street{Paul Henri Spaaklaan 1}, \postcode{6229 EN}  \city{Maastricht}, \country{The Netherlands}}}



\abstract{Phylogenies are commonly used to represent the evolutionary relationships between species, and often these phylogenies are equipped with edge lengths that indicate degrees of evolutionary difference. Given such a phylogeny, a popular measure for the biodiversity of a subset of the species is the phylogenetic diversity (PD).
But if we want to focus conservation efforts on particular species, we may use a phylogenetic diversity index, a function that shares out the PD value of an entire phylogeny across all of its species. With these indices, various species-level conservation strategies can be evaluated. 
This work explores how the most suitable diversity indices can be found. In particular, how formalizing the requirement for diversity indices to capture high levels of PD, or to maintain a scoring of taxa in the presence of uncertain edge lengths drives the selection of a suitable index. Furthermore, we provide illustrations of these new mechanisms for diversity index selection in a case study. This analysis includes the comparison to popular phylogenetic indices from the conservation literature. }




\maketitle

\section{Introduction}\label{intro}
The study of biodiversity has played a prominent role in conservation biology for centuries~\cite{myers89,Singh02,spicer13}. The current threat of extinction in many populations~\cite{iucn06} has elevated the search for prioritization protocols to protect species with high levels of evolutionary distinctiveness, genetic information or community differentiation~\cite{wright90,crozier97,hartmann06,tucker17}. In this article, we consider quantitative, phylogeny-based methods that can inform such priorities. The need for such methods is great because the scale of the conservation problem is large compared to the funding currently available for direct interventions (see~\cite{HaywardCastley2018} and citations within), though more optimistic views also exist~\cite{wiedenfeld2021conservation}.
Despite the latter perspective, until funding allocations do change there remains an economic question of where to use available resources most effectively to preserve biodiversity.

To be useful, any quantitative methods employed need to be accurate, informative and remain so as phylogenetic trees are updated. 
Without these qualities, prioritization protocols run the risk of being unhelpful or being subject to large changes when extinctions occur or when further resolution is added to a phylogeny. 
Such views have spurred recent research into the mathematical properties of existing measures~\cite{semple23, fischer23, DIopt1, manson24, moulton2024phylogenetic}. 
Another motivator is that these methods have been used as part of real-world conservation decision making processes for determining funding allocations, such as those undertaken by the EDGE of Existence programme~\cite{EDGEofExistence_2017}.
We consider two types, namely phylogenetic diversity (PD)~\cite{faith92} and phylogenetic diversity indices~\cite{DIopt1}. The former measures the evolutionary history shared by species. The latter derive from a desire to individualize phylogenetic diversity by seeking to share out or attribute the biodiversity represented by a phylogenetic tree to the individual species it contains. A key point of the present work is the tension between these differing views of what is meant by \textit{biodiversity}. 

One peculiarity of this field is that various quantitative methods appear on an ad hoc basis, as some fairly intuitive solution to a particular biological question about a particular dataset. 
The tendency for new methods to be invented, ahead of using existing ones, has lead to a so-called `jungle of indices' for biologists to choose from. 
Some have argued that the proliferation hinders rather than helps conservation management, understanding and evaluation \cite{ricotta2005through}.
Resolving this `jungle' has become a topic of interest for those interested in phylogenetic trees from a theoretical standpoint.

The main strategy has been to start with known measures and determine their properties. 
From this, some measures are rejected as lacking desirable properties (such as Faith's refutation of a method by Altshul and Lipman~\cite{faith1994biodiversity}).
Others are bolstered by having their desirable properties proved rigorously, or shown to be unique (for example, in~\cite{DIopt1}). 
However, this approach suffers by being only able to evaluate known measures (or further ad hoc inventions). 
It may well be that the optimal solution methods for phylogenetic problems are currently unknown and will remain unknown if we continue to follow only this line of research.

For this reason we instead advocate applying a `properties first' strategy to the family of phylogenetic diversity indices. This parallels an approach previously used to evaluate abundance-based diversity measures, such as Hill numbers~\cite{daly2018ecological}.
Under this strategy, the desirable properties of quantitative phylogenetic methods are established first. 
Next, a careful analysis of the properties is used to reveal the methods that give optimal solutions.
This shifts the focus on to the strength or importance of the desirable properties, while remaining agnostic about the particular methods that arise.
We expect that this approach will offer new solutions, from which practitioners will be able to select measures that better fit the properties they require. 

The present article follows the properties first strategy for two phylogenetic diversity index properties, which we outline in Sections~\ref{secDiversityDifference} and~\ref{secContinuity}. Our analysis of different indices uses Operations Research techniques to find optimal solutions to a collection of problems. 
Section~\ref{sec:1} includes an exact formulation of these problems and the key definitions of diversity indices and the set of diversity index scores. In Section~\ref{secSolution} we include mathematical results that ensure our robust optimization techniques will work properly and subsequently show that our problems can be solved in polynomial time. Next, Section~\ref{secExp} compares the optimal solutions to our problems against other popular diversity indices in the literature. Finally, Section~\ref{secRemarks} contextualizes our results and discusses further problems for which the present approach could prove fruitful.

\subsection{The diversity difference}\label{secDiversityDifference}

To motivate our investigations in this article, we consider the following conservation scenario. Suppose that we have been asked to select a given number of species $\kappa$, say, to prioritize for extra conservation funding. These $\kappa$ species must be chosen with reference to their placement in a larger phylogeny of interest. This phylogeny is assumed to be known and includes branch lengths. 
There are two ways in which to choose our species. The first is to compare leaf sets of size $\kappa$, measuring the PD of each set. This is arguably the best measure of the biodiversity of each set  \textit{as a set}~\cite{semple23}. However, for time-based phylogenies, PD does not directly indicate the contribution of individual species towards this diversity quantity. Understanding of individual contributions is more readily achieved by using a phylogenetic diversity index. These indices also can be viewed as measuring the evolutionary isolation of each species. 

These two measures approach diversity from different standpoints, leading to a trade-off, highlighted by Redding and colleagues~\cite{redding2008evolutionarily}, ``between prioritizing the most evolutionary distinctive species in a tree and prioritizing sets of species that best represent the whole tree''. 
The size of this trade-off is measured by the \textit{diversity difference} (defined in full in the next section). In this article we develop a selection mechanism for phylogenetic diversity indices that minimizes how large the diversity difference can get. Mitigating larger differences is important because the diversity difference tends to get larger the more spread out sets of species are and it is those sets that capture more PD than closely arranged sets of species do.
Because of this, we should be using diversity indices that capture high levels of PD and guarantee a lower bound on PD capture independent of any particular subset of species. A key observation made here is that this problem is not solved by well-known diversity indices from the literature.

\subsection{The diversity index continuity property}\label{secContinuity}

Next, we incorporate the \textit{diversity index continuity property}~\cite{steel16} in our analysis. This property states that the measurements made by a phylogenetic diversity index are not impacted by edge contractions and expansions.

It is not uncommon for trees containing polytomies to be resolved to binary trees by using very short edges. Yet, the uncertainty in the complete resolution of a phylogenetic tree should not have an outsized impact on the conservation decisions. We would prefer if small shifts in tree topology lead to steady changes in diversity index values, without any large jumps. Therefore, we quantify the degree of discontinuity of all diversity indices to order their sensitivity to changes in edge lengths in an underlying phylogenetic tree. Under the assumption that a fixed subset of species should be prioritized, we provide evidence that long edges determine changes in a ranking based on diversity index scores when small changes in the tree shape occur anywhere. Moreover, we assess how far solutions to the diversity difference problem deviate from the continuity property.

\section{Robust biodiversity problems}\label{sec:1}
\begin{figure}[!t]
\centering
\includegraphics[scale=0.45]{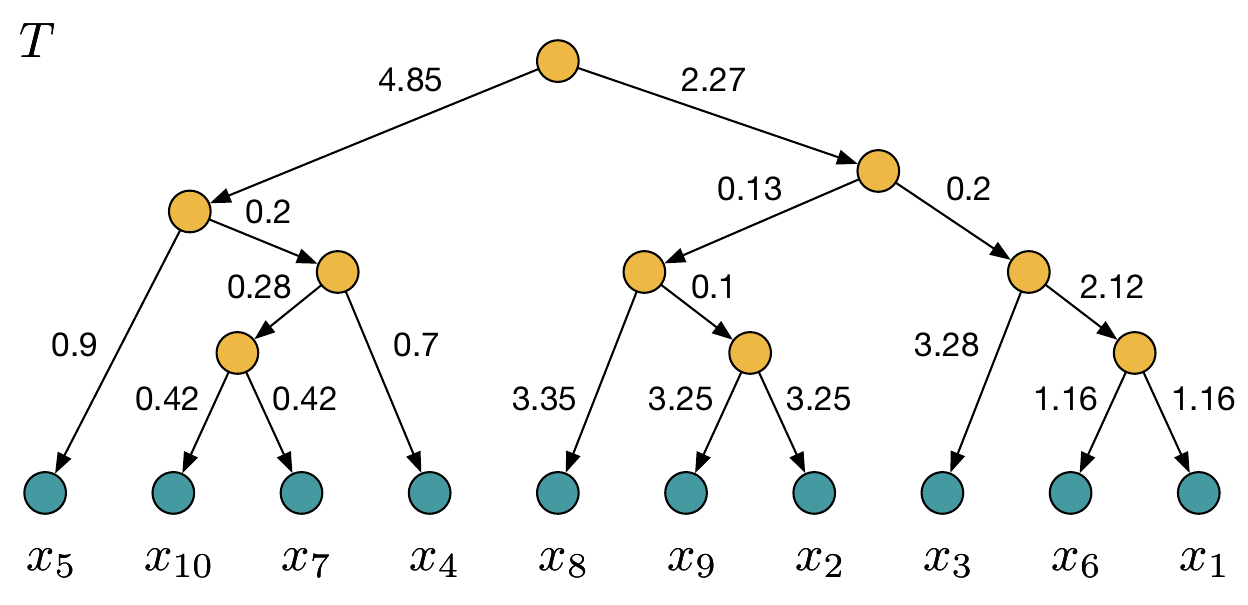}
\caption{A rooted phylogenetic $X$-tree for $X=\{x_1,x_2,\dots,x_{10}\}$ and edge lengths as shown.}\label{fig::simTree1}
\end{figure}

In this section we formalize our approach to measure and preserve biodiversity. To this end, we introduce the necessary mathematical notation and background information on diversity indices. Given a set of $n$ species $X=\{x_1,\dots,x_n\}$, called \emph{taxa}, a \emph{rooted $X$-tree} is a connected acyclic digraph $T=(V,E)$ with leafset $X$ and a vertex $\rho\in V$ with in-degree~$0$ and out-degree~$2$ such that there exists a (unique) directed path from $\rho$ to every leaf of $T$. We call $X$ the set of \emph{taxa} and $\rho$ the \emph{root} of $T$. Note that our attention shall be restricted to \emph{binary} rooted $X$-trees throughout this work, where every vertex in $V\setminus(X\cup\{\rho\})$ has in-degree~$1$ and out-degree~$2$. We shall restrict our attention to binary rooted $X$-trees throughout. For $Y\subseteq X$, let $T_Y=(V_Y,E_Y)$ denote the spanning subtree of $T$ rooted at $\rho$ with leafset $Y$ and let $\ell :E\to\mathbb{R}_{\geq 0}$ be an edge length function. For $v\in V$, let $T[v]$ denote the maximal induced subtree of $T$ rooted in $v$, called a \emph{pendant subtree} of $T$, and let $X[v]$ be the leafset of $T[v]$.  For a binary rooted $X$-tree $T$, we call the tuple $(T,\ell)$ a \emph{rooted phylogenetic $X$-tree}, an example of which is shown in Figure~\ref{fig::simTree1}. Now, we can define our first measure of biodiversity~\cite{faith92}:

\begin{definition}
Let $(T,\ell)$ be a rooted phylogenetic $X$-tree and let $Y\subseteq X$. Then, we call
\begin{align}\label{PDY}
\text{PD}_{(T,\ell)}(Y)=\sum_{e\in E_Y}\ell(e)
\end{align}
the \emph{phylogenetic diversity} of the set $Y$. 
\end{definition}

To illustrate phylogenetic diversity, consider the rooted phylogenetic $X$-tree in Figure~\ref{fig::simTree1} and $Y=\{x_2,x_3,x_8,x_9\}$. Then, we have
\begin{align*}
\text{PD}_{(T,\ell)}(Y)=2.27+0.13+3.35+0.1+2\cdot 3.25+0.2+3.28=15.83
\end{align*}
If the chosen taxa are more ``spread out'' than the taxa in $Y$, then the phylogenetic diversity increases. For example, for $Y^\prime =\{x_2,x_3,x_8,x_{10}\}$, we get
\begin{align*}
\text{PD}_{(T,\ell)}(Y^\prime)=\,\text{PD}_{(T,\ell)}(Y)-3.25+4.85+0.2+0.28+0.42=18.33.
\end{align*}
To define our second measure of biodiversity, which apportions the biodiversity of $X$ across a subset of species, we consider a weighted version of phylogenetic diversity~\cite{DIopt1}. Specifically, for a rooted phylogenetic $X$-tree $\hat{T}=(T,\ell)$, we call the linear function $\varphi_{\hat{T}}:X\to\mathbb{R}_{\geq 0}^{n}$ defined by
\begin{align}\label{DIfunction}
\varphi_{\hat{T}}(x)=\sum_{e\in E}\gamma(x,e)\ell(e)
\end{align}
a \emph{diversity index} of $\hat{T}$ where, for $x\in X$, $e\in E$, the $\gamma(x,e)$ coefficients are non-negative real numbers satisfying the following conditions:
\begin{enumerate}
\item (Convexity) $\sum_{x\in X}\varphi_{\hat{T}}(x)=\,\text{PD}_{\hat{T}}(X)$.
\item (Descent) $\gamma(x,(u,v))=0$ if there exists no directed path in $T$ from $v$ to $x$.
\item (Neutrality) $\gamma(x,(u,v))$ is a function of the tree shape of $T[v]$. In addition, for $e_i=(u_i,v_i)\in E$, $i\in\{1,2\}$, if there exists an isomorphism $\phi$ from $T[v_1]$ to $T[v_2]$, then $\gamma(x,e_1)=\gamma(\phi(x),e_2)$ for all $x\in X[v_1]$.
\end{enumerate}
For example, if we define function~\eqref{DIfunction} such that, for all $x\in X$, $(u,v)\in E$, we choose coefficients $\gamma(x,(u,v))$ for which $-\log_2(\gamma(x,(u,v)))$ is equal to the number of edges on the path in $T$ from $v$ to $x$, then conditions~(1) to~(3) are satisfied. Here, conditions (2) and (3) hold by definition. To see that the convexity condition holds, let $m_{vx}$ denote the number of edges on the directed path in $T$ from $v\in V(T)$ to $x\in X$. Then, we can write $\gamma(x,(u,v))=2^{-m_{vx}}$ for all $x\in X$, $(u,v)\in E$. Hence, 
\begin{align*}
\sum_{x\in X}\varphi_{\hat{T}}(x)&=\sum_{x\in X}\sum_{e=(u,v)\in E}\gamma(x,(u,v))\cdot\ell(e)\\
&=\sum_{e=(u,v)\in E}\sum_{x\in X[v]}2^{-m_{vx}}\cdot\ell(e)\stackrel{T[v]\text{ binary}}{=}\sum_{e\in E}\ell(e)=\,\text{PD}_{\hat{T}}(X).
\end{align*}
This index is called the \emph{Equal Splits Index} (ES) in the literature~\cite{redding06}. We denote the set of images of diversity indices $\varphi_{\hat{T}}$ by $\mathcal{P}(T,\ell)$ and define the measurement of a diversity index with respect to a subset of taxa as follows:

\begin{definition}
Let $(T,\ell)$ be a rooted $X$-tree, let $s\in\mathcal{P}(T,\ell)$ and let $Y\subseteq X$. Then, we call
\begin{align*}
\text{ID}_{(T,\ell)}(Y,s)=\sum_{x_i\in Y}s_i
\end{align*}
the \emph{indexed diversity} of $Y$ and $s$.
\end{definition}

Observe that the phylogenetic diversity of $X$ is an upper bound for the indexed diversity of $Y\subseteq X$ because of the convexity condition~(1). Moreover, the phylogenetic diversity of any pendant subtree of $(T,\ell)$ is a lower bound for the indexed diversity of its leafset in $(T,\ell)$. To illustrate this observation, consider the rooted phylogenetic $X$-tree $(T,\ell)$ and set of taxa $Y=\{x_4,x_7,x_{10}\}$ in Figure~\ref{fig::simTree1}. Then, the pendant subtree $T[v]$ of $T$ with leafset $X[v]=Y$ yields $$\text{PD}_{(T[v],\ell)}(Y)=0.28+2\cdot 0.42+0.7=1.82.$$ Hence, for any $s\in\mathcal{P}(T,\ell)$ and $s'\in\mathcal{P}(T[v],\ell)$, 
\begin{align*}
\text{ID}_{(T,\ell)}(Y,s)\geq\,\text{ID}_{(T[v],\ell)}(Y,s')\stackrel{\text{(Convexity)}}{=}\,\text{PD}_{(T[v],\ell)}(Y)=1.82.
\end{align*}
Observing these bounds gives rise to the question of just how much the PD and ID values differ across sets of leaves. In other words, we are interested in the function $\Delta_{\hat{T}}:2^X\times\mathcal{P}(T,\ell)\to\mathbb{R}_{\geq 0}$ defined by
\begin{align}\label{def::Delta}
\Delta_{\hat{T}}(Y,s)=\,\text{PD}_{\hat{T}}(Y)-\,\text{ID}_{\hat{T}}(Y,s).
\end{align}
For example, suppose we choose ES as our index on the tree $\hat{T}$ in Figure~\ref{fig::simTree1} to obtain a diversity index score $s^{\text{ES}}$. Then, for $Y=\{x_1,x_2,x_3,x_4\}$ and $Z=\{x_4,x_5,x_7,x_{10}\}$,
\begin{align*}
\Delta_{\hat{T}}(Y,s^{\text{ES}})= 18.26 - (2.555 + 3.6175 + 3.94 + 2.0125) = 6.135
\end{align*}
and
\begin{align*}
\Delta_{\hat{T}}(Z,s^{\text{ES}})= 7.77 - (2.0125 + 3.325 + 1.21625 + 1.21625) = 0.
\end{align*}
This means ES captures more PD for taxa $Z$ than for taxa $Y$. However, we also see that taxa $Z$ are vastly more ``spread out'' than taxa $Y$ such that the increase in PD is significantly larger than the increase in ID. This relationship between $\Delta_{\hat{T}}(Y,s)$ and visual spread is typical. In this article we analyze the ability of diversity indices to capture PD without requiring any particular structure of the underlying taxa set. Then, the deviation between PD and ID can be quantified for any fixed number of species (denoted by $\kappa$) as follows:
\begin{problem}\label{prob::MDDP}
Let $\hat{T}=(T,\ell)$ be a rooted phylogenetic $X$-tree, let $s\in\mathcal{P}(T,\ell)$ and let $\kappa\geq 2$ be an integer. Then, we call
\begin{align*}
\text{MDD}_{\hat{T}}(s,\kappa)=\max_{Y\subseteq X,\,|Y|=\kappa}\Delta_{\hat{T}}(Y,s)
\end{align*}
the \emph{Maximum Diversity Difference Problem (MDDP)} of $s$ and order $\kappa$.
\end{problem}
Now, for each $Y\subseteq X$, $|Y|=\kappa$, let $\varphi_{\hat{T}}^Y$ denote the index (with image $s^Y$) which minimizes $\Delta_{\hat{T}}(Y,s^Y)$, i.e., $\varphi_{\hat{T}}^Y$ captures high levels of PD in $\hat{T}$. Then, we call $s\in\mathcal{P}(T,\ell)$ such that
\begin{align}\label{ineq::robustBound}
\text{MDD}_{\hat{T}}(s,\kappa)&\leq\max_{Y\subseteq X,\,|Y|=\kappa}\Delta_{\hat{T}}(Y,s^Y)
\end{align}
the image of a \emph{robust diversity index}. This means, any index from the family of robust diversity indices captures high levels of PD and guarantees a lower bound on PD capture independent of the choice of taxa by imposing upper bound~\eqref{ineq::robustBound}. Moreover, we solve the MDDP with no expectation of which diversity index should be used. Indeed, the right-hand side of bound~\eqref{ineq::robustBound} is independent of the choice of $s$. Hence, robust diversity indices quantify the deviation between PD and ID using set $\mathcal{P}(T,\ell)$ as a region of uncertainty. In the language of robust optimization we identify the worst case for the minimum difference between PD and ID across the different scenarios $s\in \mathcal{P}(T,\ell)$~\cite{yu2013}:
\begin{problem}\label{prob::RDDP}
Let $\hat{T}=(T,\ell)$ be a rooted phylogenetic $X$-tree and let $\kappa\geq 2$ be an integer. Then, we call
\begin{align*}
\text{RDD}_{\hat{T}}(\kappa)=\max_{Y\subseteq X,\,|Y|=\kappa}\min_{s\in \mathcal{P}(T,\ell)}\Delta_{\hat{T}}(Y,s)
\end{align*}
the \emph{Robust Diversity Difference Problem (RDDP)} of order $\kappa$.
\end{problem}
Here, the name of the RDDP comes from the fact that any optimal solution to the RDDP is the image of a robust diversity index. Despite the similarity between the RDDP and MDDP it's not automatically clear how to solve the RDDP. The MDDP can be solved in polynomial-time by a greedy algorithm~\citep{semple23}. This greedy algorithm relies on the fact that the non-negative function~\eqref{def::Delta} is linear in $Y$ for constant $s$. Hence, it cannot be directly adapted to solve the RDDP. Furthermore, notice that the definition of Problem~\ref{prob::RDDP} is very restrictive because we require robustness for the whole uncertainty region $\mathcal{P}(T,\ell)$. That is, upper bound~\eqref{ineq::robustBound} might be very small because for every set of species there can exist a diversity index which captures high levels of PD. If such restrictive indices exist only in the margins of region $\mathcal{P}(T,\ell)$, they can in theory deprive robust diversity indices of other desirable properties or make optima of the RDDP unappealing from a practical point of view. Therefore, to alleviate restrictions on the region of robust diversity indices, we will also consider relaxations of the RDDP in which we minimize only over a proper subset of $\mathcal{P}(T,\ell)$.\\~\\
Finally, we formalize our study of continuity among diversity indices. This aspect concerns the sensitivity of the descent and neutrality conditions in the definition of diversity indices to changes in the shape of the rooted phylogenetic $X$-tree $(T,\ell)$. Small changes in $(T,\ell)$ ideally correspond to no or only small changes in diversity index scores. To make use of this desirable property, we quantify diversity indices with different levels of sensitivity to changes in the tree shape $T$. Specifically, we consider the shape of $T$ as a region of uncertainty in which we allow for the collapsing of individual edges. We denote $T/e$ as the rooted $X$-tree we obtain from $T$ by collapsing edge $e\in E$. Conversely, we could study the expansion of an unresolved rooted $X$-tree as a region of uncertainty by adding edges. Since we assume our given tree to be binary, we stick to the former definition of an uncertain tree shape. Then, a diversity index $\varphi_{(T,\ell)}$ which satisfies
\begin{align*}
\lim_{\ell(e)\to 0}\varphi_{(T,\ell)}(x)=\varphi_{(T\setminus e,\ell)}(x)~&~&\forall\,x\in X,\,e\in E
\end{align*}
is called \emph{continuous}~\cite{steel16}. Any continuous diversity index would be ideal given these considerations for uncertain tree shapes. Unfortunately, only one continuous diversity index exists (Theorem 6 in~\citep{DIopt1}):

\begin{proposition}\label{prop::cont1}
Let $\hat{T}=(T,\ell)$ be a rooted phylogenetic $X$-tree and let $\varphi_{\hat{T}}$ be a diversity index of $\hat{T}$. Then, $\varphi_{\hat{T}}$ is continuous if and only if
\begin{align}\label{def::cont1}
\gamma(x,(u,v))=\frac{1}{|X[v]|}~&~&\forall\,(u,v)\in E,\,x\in X[v].
\end{align}
\end{proposition}

The diversity index satisfying equations~\eqref{def::cont1} is called the \emph{Fair Proportion Index}~(FP) in the literature~\cite{hartmann13}. We want to relax the definition of continuity to obtain more diversity indices which are (locally) robust with respect to an uncertain tree shape. To this end, we consider the following reformulation of Proposition~\ref{prop::cont1} which follows directly from the convexity condition on diversity indices:

\begin{corollary}\label{cor::cont2}
Let $\hat{T}=(T,\ell)$ be a rooted phylogenetic $X$-tree and let $\varphi_{\hat{T}}$ be a diversity index of $\hat{T}$.  Then, $\varphi_{\hat{T}}$ is continuous if and only if coefficients $\gamma(x,e)$ take on exactly one non-zero value for all $e\in E$.
\end{corollary}

Now, we relax continuity as characterized by Corollary~\ref{cor::cont2} by considering the dependence of coefficients $\gamma(x,e)$ on $e\in E$ \emph{and} $x\in X$ and measuring the deviation from the continuity assumption. 
\begin{definition}
Let $\hat{T}=(T,\ell)$ be a rooted phylogenetic $X$-tree and let $\varphi_{\hat{T}}$ be a diversity index of $\hat{T}$. For a constant $\theta\in [0,1]$, we call a diversity index $\varphi_{\hat{T}}$\linebreak \emph{$\theta$-compatible} if and only if there exists $\delta\in\mathbb{R}_{\geq 0}^n$ such that
\begin{align}
\delta_i&\geq\left[\gamma(x_i,e)-\gamma(x_j,e)\right]\ell(e)~&~&\forall\,e=(u,v)\in E,\,x_i,x_j\in X[v],\,i\neq j,\label{comp::con1}\\
\sum_{x_i\in X}\delta_i&\leq(1-\theta)\cdot\,\text{PD}_{(T,\ell)}(X).\label{comp::con3}
\end{align}
\end{definition}
Observe that $\varphi_{\hat{T}}$ is continuous if and only if $\varphi_{\hat{T}}$ is $1$-compatible. Moreover, if $\theta=0$, we know that the right-hand side of inequality~\eqref{comp::con3} provides the least upper bound on the sum of entries in $\delta$ which satisfy conditions~\eqref{comp::con1} and~\eqref{comp::con3} for $0$-compatibility. Indeed, all diversity indices are $0$-compatible because in this case inequality~\eqref{comp::con3} holds with equality only if there exists a bijection $\phi:X\to E$ such that $\gamma(x_i,\phi(x_i))\ell(\phi(x_i))$ is maximum among all diversity indices. This also means that for any $\theta\in(0,1]$ at least one diversity index is not $\theta$-compatible. 
Let $\mathcal{P}(T,\ell,\theta)$ denote the set of tuples of diversity indices scores corresponding to $\theta$-compatible diversity indices of $\hat{T}$ and vectors $\delta$ satisfying constraints~\eqref{comp::con1} and~\eqref{comp::con3}. Then, the restriction to $\theta$-compatible diversity indices does not change PD but has an effect on ID for a sufficiently small number $\epsilon>0$:

\begin{problem}\label{prob::MCDP}
Let $\hat{T}=(T,\ell)$ be a rooted phylogenetic $X$-tree and let $\kappa\geq 2$, $\theta\in[0,1]$ be integers. Then, we call
\begin{align*}
\text{MCD}_{(T,\ell)}(\kappa,\theta)=\max_{\substack{Y\subseteq X,\,|Y|=\kappa\\ (s,\delta)\in\mathcal{P}(T,\ell,\theta)}}\text{ID}_{(T,\ell)}(Y,s)-\epsilon\cdot\sum_{x_i\in X}\delta_i
\end{align*}
the \emph{Maximum Compatible Diversity Problem} (MCDP) of order $(\kappa,\theta)$.
\end{problem}

The purpose of the penalty term in the objective function of the MCDP depending on $\epsilon$ is to allow for comparisons between optimal solutions to the MCDP with the same objective function value for $\epsilon=0$ but different input tree $\hat{T}$. Since we want to compare optimal solutions to the MCDP for small changes in $\hat{T}$, this property will be important to keep changes in the left-hand sides of constraints~\eqref{comp::con1} and~\eqref{comp::con3} to a minimum. Specifically, to measure the change in ID when an edge is collapsed we are interested to calculate and compare $\text{MCD}_{(T,\ell)}(\kappa,\theta)$ and $\text{MCD}_{(T\setminus e,\ell)}(\kappa,\theta)$. Let $s^*$ and $s^e$ denote the respective optimal solutions. Furthermore, let $\varphi_{(T,\ell)}$ and $\varphi_{(T\setminus e,\ell)}$ be the diversity indices with image $s^*$ and $s^e$, respectively. Then, we measure the change in diversity index coefficients when collapsing edge $e$ by calculating the \emph{symmetrized Kullback-Leibler divergence}~\cite{khalid17}
\begin{align*}
D_{\text{KL}}^{\text{sym}}(p_j||q_j)=\frac{1}{2}\cdot\sum_{e_j=(u_j,v_j),\,x_i\in X[v_j]}\left[p_{ij}\cdot\ln\left(\frac{p_{ij}}{q_{ij}}\right)+q_{ij}\cdot\ln\left(\frac{q_{ij}}{p_{ij}}\right)\right]
\end{align*}
for the probability distributions $p_j$ and $q_j$ of coefficients $\gamma(x_i,e_j)$, $e_j=(u_j,v_j)\in E$, $x_i\in X[v_j]$ for diversity indices $\varphi_{(T,\ell)}$ and $\varphi_{(T\setminus e,\ell)}$, respectively. Furthermore, we call
\begin{align*}
\text{dd}(T,\ell,\kappa,\theta,e)=\sum_{e_j\in E(T\setminus e)}D_{\text{KL}}^{\text{sym}}(p_j||q_j)
\end{align*}
the \emph{degree of discontinuity} of $\hat{T}$ with respect to order $(\kappa,\theta)$ and edge $e$.

Observe that $D_{\text{KL}}^{\text{sym}}(p_j||q_j)$ is an information-theoretic function measuring a distance between probability distributions $p_j$ and $q_j$. Hence, the degree of discontinuity of $\hat{T}$ is a distance function depending on all coefficients in $\varphi_{(T,\ell)}$ and $\varphi_{(T\setminus e,\ell)}$ defined for edges in $E(T\setminus e)$. For example, $\text{dd}(T,\ell,n,\theta,e)=0$ because of the convexity condition of diversity indices and $\text{dd}(T,\ell,\kappa,1,e)=0$ because FP is the unique $1$-compatible diversity index. This concludes our formalization of robust diversity indices and diversity index selection under uncertain tree shapes.

\section{Exact solution algorithms for the RDDP and MCDP}\label{secSolution}
In this section we show that the RDDP and MCDP can be solved in polynomial time. To this end, we utilize a particular description of the space of diversity indices $\mathcal{P}(T,\ell)$. By defining upper and lower bounds on diversity index scores for each taxon, the space can be viewed in such a way to make it amenable to the techniques of convex optimization. An example follows the results in the subsection below, and further details underpinning this approach can be found in a companion paper~\cite{frohn24}.

\subsection{Combinatorial properties of diversity indices}\label{sec2}
We begin by extending the notation of the previous section. Let $T$ be a rooted $X$-tree. We write $V_{\text{int}}(T)$, $V(T)$ and $E(T)$ to refer to the set of interior vertices, all vertices and all edges of $T$, respectively. For an edge length function $\ell :E(T)\to\mathbb{R}_{\geq 0}$, to simplify our notation, we denote any restriction of $\ell$ to subsets of $E(T)$ by $\ell$, too. We denote the directed path in $T$ from $\rho$ to a vertex $v\in V(T)$ by $P(v)$, and the path between vertices $u,v\in V(T)$ by $P(u,v)$. We call $T$ \emph{balanced} if $P(x)$ has the same length for all $x\in X$. We call a diversity index $\varphi_{\hat{T}}$ \emph{consistent} if and only if for all $v\in V_{\text{int}}(T)$, $e_1,e_2\in P(v)$ there exists $k_v\in\mathbb{R}_{\geq 0}$ such that
\begin{align*}
\sum_{x\in X[w]}\gamma(x,e_1)&=k_v\cdot\sum_{x\in X[w]}\gamma(x,e_2)~&~&\forall\,(v,w)\in E(T).
\end{align*}
We denote $d_{(T,\ell)}$ as the dimension of $\mathcal{P}(T,\ell)$ when requiring all diversity indices to be consistent. For $Z\subseteq Y\subseteq X$, let $F(Y-Z)$ denote the forest induced by removing all directed paths from the root of $T$ to taxa in $Z$ from $T_Y$. Furthermore, for $x_k\in Y$, let
\begin{align*}
\mathcal{Z}(x_k,Y)=\left\{Z\subseteq Y\setminus\{x_k\}\,:\,Z\neq\emptyset,~F(Y-Z)\text{ is a tree}\right\}.
\end{align*}
Then, we summarize some results on diversity indices that will be useful in this article:
\begin{proposition}\label{prop::MS}
(\citet{DIopt1}) Let $\hat{T}=(T,\ell)$ be a rooted phylogenetic $X$-tree. Then $\mathcal{P}(T,\ell)$ is a polytope and, for $x\in X$, there exist non-negative real numbers $\text{LB}(x,T)$ and $\text{UB}(x,T)$ such that $\text{LB}(x,T)\leq\varphi_{\hat{T}}(x)\leq\,\text{UB}(x,T)$ for all diversity indices~$\varphi_{\hat{T}}$.
\end{proposition}
\begin{proposition}\label{prop::FM}
(\citet{frohn24}) Let $\hat{T}=(T,\ell)$ be a rooted phylogenetic $X$-tree with $d_{\hat{T}}\geq 1$. For $s\in\mathcal{P}(T,\ell)$, $x_i\in X$, inequalities 
\begin{align}\label{facet1}
s_i\geq\,\text{LB}\left(x_i,T\right)
\end{align}
are facet-defining for $\mathcal{P}(T,\ell)$. Moreover, for $x_i\in X$, there exist sets $Y_i\subseteq X$ with $|Y_i|=d_{\hat{T}}+1$, $x_k\in Y$, $k\neq i$, real numbers $N_j$ and $r_Z$ such that inequalities
\begin{align}\label{facet2}
\sum_{x_j\in Z}N_j\cdot\left(s_j-\,\text{LB}\left(x_j,T\right)\right)&\leq r_Z~&~&\forall\,Z\in\mathcal{Z}\left(x_k,Y_i\right),\,x_i\in Z
\end{align}
are facet-defining for $\mathcal{P}(T,\ell)$. In addition, facets of form~\eqref{facet1} and~\eqref{facet2} constitute a minimal compact description of $\mathcal{P}(T,\ell)$.
\end{proposition}
\begin{figure}[!t]
\centering
\includegraphics[scale=0.45]{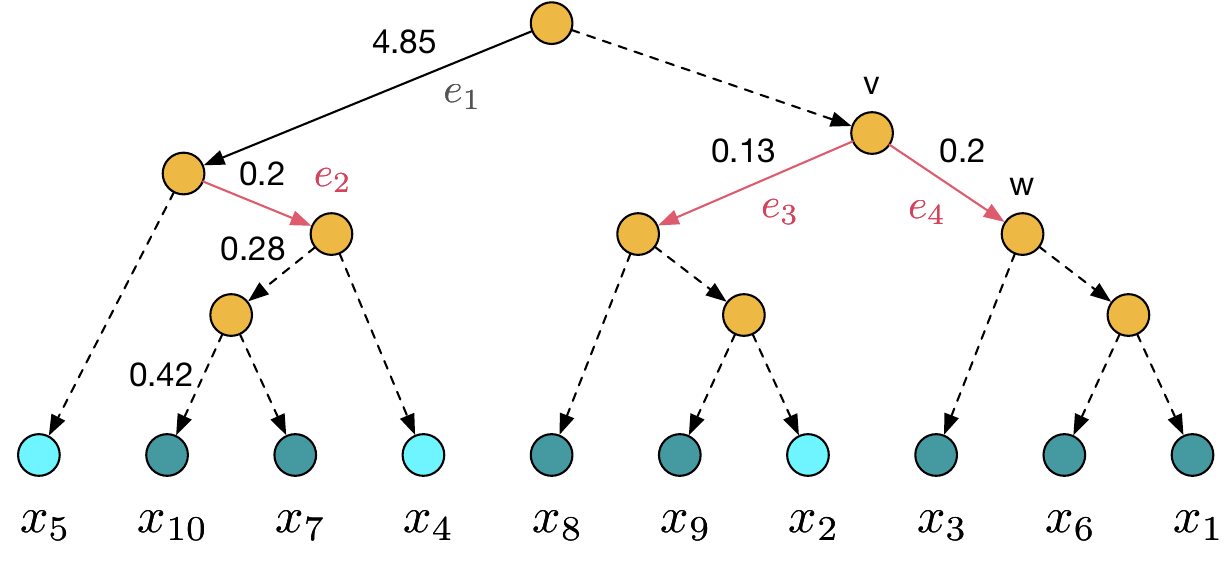}
\caption{The rooted phylogenetic $X$-tree from Figure~\ref{fig::simTree1}. Dashed edges $e$ indicate that coefficients $\gamma(x,e)$, $x\in X$, in~\eqref{DIfunction} are identical across all consistent diversity indices. Red edges $\{e_2,e_3,e_4\}$ share a neutrality condition. The light blue taxa form a set $Y_5$ to which Proposition~\ref{prop::FM} applies.}\label{fig::simTree2}
\end{figure}
We illustrate the construction of some such lower and upper bounds in the following example:
\begin{example}\label{exIntro}
Consider the rooted phylogenetic $X$-tree in Figure~\ref{fig::simTree2}: edges $(v,x)$, $x\in X$, called \emph{pendant edges}, satisfy $\gamma(x,(v,x))=1$ due to the convexity and descent conditions. Dashed edges which are not pendant have coefficients that must satisfy neutrality and consistency conditions. Here, consistency conditions can be viewed as localizing the choice of coefficients $\gamma(x,(u,v))$ to vertices $v$. In particular, if $T[v]$ is balanced up to replacements of a fixed subtree shape by a singleton, then consistency conditions imply that coefficients $\gamma(x,(u,v))$, $x\in X[v]$, are identical for all consistent diversity indices. This means, for vertices $v$ and $w$ in Figure~\ref{fig::simTree2}, replacing all subtrees $T[u]$, $u\in V(T)$, with the same shape as $T[w]$ by a singleton transforms $T[v]$ into a balanced tree. From these observations one can derive the lower and upper bounds in Proposition~\ref{prop::MS}. For example, from the edge lengths in the path $P(x_{10})$ we conclude that
\begin{align*}
\text{LB}\left(x_{10},T\right)&=0.42+0.5\cdot 0.28=0.56,\\
\text{UB}\left(x_{10},T\right)&=\,\text{LB}\left(x_{10},T\right)+0.2+4.85=5.61.
\end{align*}
\end{example}
For details on the construction of sets $Y_i$ in Proposition~\ref{prop::FM}, see Proposition~3.2 in~\cite{frohn24}. Here, we just note that $Y_i$ is not unique and depends on a choice of representatives for equivalence classes induced by the neutrality conditions. Indeed, a bijection between edges $(a,b)\in\{e_1,\dots,e_4\}$ in Figure~\ref{fig::simTree2} to taxa $X[b]\cap Y_i\setminus\{x_i\}$ is the basis of this construction. 
In addition, we note that numbers $N_j$ and $r_Z$ in Proposition~\ref{prop::FM} can be calculated in linear time~\cite{frohn24}.

In view of the RDDP, we see that the description of space $\mathcal{P}(T,\ell)$ by Propositions~\ref{prop::MS} and~\ref{prop::FM} provides a model which allows us to solve the RDDP in variables $s\in\mathcal{P}(T,\ell)$. However, when considering the combinatorial aspects of the MCDP we require constraints~\eqref{comp::con1} which can not always be expressed in variables $s\in\mathcal{P}(T,\ell)$. For example, if a vertex $v$ is incident to three edges $e_1=(v,w_1)$, $e_2=(v,w_2)$ and $e_3=(u,v)$ for which non-zero coefficients $\gamma(x,e_i)$, $i\in\{1,2,3\}$, $x\in X$, are not constant across all diversity indices, then $d_{\hat{T}}\geq 3$ but all diversity index scores contain $\gamma(x,e_1)\ell(e_1)+\gamma(x,e_3)\ell(e_3)$ or $\gamma(x,e_2)\ell(e_2)+\gamma(x,e_3)\ell(e_3)$ or none of edges $e_i$, $i\in\{1,2,3\}$. This means, only two out of the three non-redundant constraints~\eqref{comp::con1} formed by $\gamma(x,e_i)$, $i\in\{1,2,3\}$, can be expressed as a linear function of diversity index scores. Hence, to model the MCDP we introduce variables $s_i^e$ for all edges $e\in E$ such that $s_i=\sum_{e\in E}s_i^e$ for all $s\in\mathcal{P}(T,\ell)$, $x_i\in X$, to decompose the linear function~\eqref{DIfunction} into terms $s_i^e=\gamma(x_i,e)\ell(e)$. In the next section we provide the specifics for both a model for the RDDP and the MCDP, respectively.

\subsection{On the optimization aspects of the RDDP and MCDP}\label{sec3}
In this subsection we propose exact solution algorithms for Problems~\ref{prob::RDDP} and~\ref{prob::MCDP}. First, we focus on the RDDP. Observe that function~\eqref{def::Delta} is bilinear in arguments $Y$ and $s$. This means, we know from the minimax theorem that
\begin{align*}
\text{RDD}_{\hat{T}}(\kappa)=\min_{s\in\mathcal{P}(T,\ell)}\max_{Y\subseteq X,\,|Y|=\kappa}\Delta_{\hat{T}}(Y,s).
\end{align*}
Equivalently, we can write the following formulation.
\begin{form}\label{form::RDDP}
\begin{align}
\text{RDD}_{\hat{T}}(\kappa)=\min~~~\alpha(s)~~~~~\,&\nonumber\\
\text{s.t.}~~~\Delta_{\hat{T}}(Y,s)&\leq\alpha(s)~&~&\forall\,Y\subseteq X,\,|Y|=\kappa\label{con::Y}\\
s&\in\mathcal{P}(T,\ell)\nonumber\\
\alpha(s)&\in\mathbb{R}_{\geq 0}\nonumber
\end{align}
\end{form}
As we have seen in the last section there exists a polynomial-sized minimal compact description of polytope $\mathcal{P}(T,\ell)$. Hence, we can calculate $\text{RDD}_{\hat{T}}(\kappa)$ by solving the polynomial-sized linear programming Formulation~\ref{form::RDDP}. We can improve this solution approach by observing that only diversity index scores $s_i$ for which taxa $x_i$ appear in constraint~\eqref{con::Y} are necessary to calculate $\Delta_{\hat{T}}(Y,s)$. We show that we can significantly reduce the number of decision variables and constraints by studying the following nested subproblem of the RDDP:
\begin{problem}\label{prob::MISP}
Let $\hat{T}=(T,\ell)$ be a rooted phylogenetic $X$-tree and let $Y\subseteq X$. Then, we call
\begin{align*}
\text{MIS}_{\hat{T}}(Y)=\min_{s\in\mathcal{P}(T,\ell)}\Delta_{\hat{T}}(Y,s)
\end{align*}
the \emph{Minimum Indexed Score Problem (MISP)} of $Y$. 
\end{problem}
Notice that the MDDP and the MISP share the same objective function but are complementary in their problem definition. The MDDP seeks to find a set of $\kappa$ taxa which maximizes the diversity difference of a given diversity index. In contrast, the MISP considers a fixed set of taxa $Y$ and identifies a diversity index for which $Y$ attains its minimum diversity difference. The details of how Proposition~\ref{prop::FM} can be used to solve the MISP in polynomial time by linear programming are given in Appendix~\ref{appendix:MISP}.


Now, we can determine $\text{RDD}_{\hat{T}}(\kappa)$ by solving the MISP for all $Y\subseteq X$, $|Y|=\kappa$, and picking the minimum among the resulting solutions. We will use this algorithm for the rest of the article because it involves less decision variables than required for solving Formulation~\ref{form::RDDP}. Further improvements to solve the RDDP faster need further research on the combinatorial properties of the diversity difference $\Delta_{\hat{T}}(Y,s)$ with respect to the potential selected sets $Y$ of $\kappa$ species.

The MCDP can be solved in a similar fashion to the RDDP. The only difference is the restriction of $\mathcal{P}(T,\ell)$ to the set of $\theta$-compatible diversity index scores $\mathcal{P}(T,\ell,\theta)$. This means we must include the linear inequalities~\eqref{comp::con1} and~\eqref{comp::con3} in our maximization of quantities $\text{ID}_{(T,\ell)}(Y,s)$ among all $Y\subseteq X$, $|Y|=\kappa$. To do so, consider the decomposition $s_i=\sum_{e\in E}s_i^e$ from the previous section to write inequalities~\eqref{comp::con1} as $\delta_i\geq s_i^e-s_j^e$. Thus, any linear description of a subset of $\mathcal{P}(T,\ell)$ in variables $s$ can be extended to a linear description of a subset of $\mathcal{P}(T,\ell,\theta)$ in variables $s^e$ and $\delta$.

\section{A comparative study of the robust and compatible diversity indices}\label{secExp}
In this section we apply our methods to a phylogenetic tree of albatrosses and two trees generated by a Yule process. We compare and contrast the optimal solutions to the RDDP and the MCDP on these trees with two well-known diversity indices from the literature, namely ES and FP. We recall their definitions from Section~\ref{sec:1}: for the number of edges on the directed path in $T$ from $v\in V(T)$ to $x\in X$ denoted by $m_{vx}$, the Equal Splits Index $\varphi_{\hat{T}}^{\text{ES}}$ is defined by
\begin{align*}
\varphi_{\hat{T}}^{\text{ES}}(x)&=\sum_{e=(u,v)\in E}2^{-m_{vx}}\cdot\ell(e)~&~&\forall\,x\in X.
\end{align*}
Also recall that the Fair Proportion Index $\varphi_{\hat{T}}^{\text{FP}}$ is defined by
\begin{align*}
\varphi_{\hat{T}}^{\text{FP}}(x)&=\sum_{e=(u,v)\in E}\frac{1}{\left|X[v]\right|}\cdot\ell(e)~&~&\forall\,x\in X.
\end{align*}
To compare ES and FP to the optimal solution to the RDDP we consider the MDDP for the image of ES and FP, respectively. We denote the resulting optimization problems by EDDP and FDDP, respectively. 

Also recall from Section~\ref{sec:1} that an optimal solution to the RDDP, i.e., a robust diversity index, can potentially exist on the boundary of region $\mathcal{P}(T,\ell)$. Hence, for comparative purposes we consider a relaxation of the RDDP in which we wish to avoid diversity indices that have large coefficients $\gamma(x,e)$. This may be motivated by observing that the coefficients in the definitions of $\varphi_{\hat{T}}^{\text{ES}}$ and $\varphi_{\hat{T}}^{\text{FP}}$ are at most $1/2$ except for pendant edges.
To restrict our search in this way, let $\mathcal{Q}(T,\ell)$ denote the subset of diversity index scores $s\in\mathcal{P}(T,\ell)$ which satisfy 
\begin{align}\label{con::Peps}
s_i^e&\leq\frac{1}{2}\cdot\ell(e),
\end{align}
for all $x_i \in X$ and all non-pendant edges.
We call $\mathcal{Q}(T,\ell)$ the set of \emph{dispersed diversity index scores}. Finally, let $\overline{\text{RDD}}_{\hat{T}}(\kappa)$ denote the objective function of Problem~\ref{prob::RDDP} restricted to dispersed diversity index scores. We call the resulting problem the R3DP of order $\kappa$.

The restricted region $\mathcal{Q}(T,\ell)$ helps to avoid diversity indices where the allocation of one taxon excessively dominates the others. We note that the value of $1/2$ is the most restrictive uniform choice for binary trees due to the convexity condition of diversity indices, and may be replaced by a value from the interval $(1/2,1]$ for a less stringent set of conditions.\\~\\
All of the implementations used in this section have been coded in ANSI C++, by relying on Gurobi Optimizer libraries v9.5.1 for linear programming. The codes have been compiled by the Apple clang compiler version 15.0.0 and run on a Quad-Core Intel Core i5 Processor at 2 GHz and 16 GB RAM, operating system macOS version 13.5.2 (kernel Darwin 22.6.0). All data and codes used in our comparative study can be downloaded at \url{https://github.com/mfrohn/OptDivIndex}.

\subsection{Selecting diversity indices under uncertainty}
 \begin{table*}[!b]
\centering
    \begin{tabular}{lllllllll}
     \toprule
        & \multicolumn{8}{l}{Species $\kappa$ selected in the Yule tree on $10$ taxa from Figure~\ref{fig::simTree1}}  \\ \cmidrule{2-9}
        & $2$ & $3$ & $4$ & $5$ & $6$ & $7$ & $8$ & $9$ \\ \midrule \\ [-1em]
	$\text{RDD}_{\hat{T}}(\kappa)=$ & $5.542$ & $5.172$ & $5.083$ & $4.400$ & $3.831$ & $2.671$ & $1.729$ & $1.059$ \\ \\ [-1em]
	$\overline{\text{RDD}}_{\hat{T}}(\kappa)=$ & $5.542$ & $5.420$ & $5.083$ & $4.482$ & $4.116$ & $2.902$ & $1.760$ & $1.394$ \\ \\ [-1em]
	$\text{FDD}_{\hat{T}}(\kappa)=$ & $7.012$ & $6.765$ & $6.342$ & $5.896$ & $5.424$ & $4.211$ & $2.929$ & $1.506$  \\ \\ [-1em]
	$\text{EDD}_{\hat{T}}(\kappa)=$ & $7.748$ & $7.606$ & $7.240$ & $6.606$ & $5.936$ & $5.136$ & $3.821$ & $2.427$\\ \midrule
	Time RDDP (sec.) & $0.012$ & $0.017$ & $0.026$ & $0.032$ & $0.029$ & $0.016$ & $0.006$ & $0.001$ \\
	Time R3DP (sec.) & $0.009$ & $0.014$ & $0.022$ & $0.026$ & $0.022$ & $0.014$ & $0.006$ & $0.001$ \\
	\bottomrule \\
    \end{tabular}
    \caption{The optimal objective function values for the RDDP, R3DP, FDDP and EDDP for different choices of~$\kappa$ for the phylogenetic $X$-tree $\hat{T}$ in Figure~\ref{fig::simTree1}. In addition, we note the solution time to solve the RDDP and R3DP of order $\kappa$ in seconds.}
    \label{tab::fig1}
\end{table*}
We initiate our comparative analysis by looking at the RDDP, R3DP, EDDP and FDDP and considering simulated input instances. Our first examples use phylogenetic trees generated by a Yule process, using the \emph{Treesim R} package~\cite{stadler11}. The Yule process is a simple pure-birth model and begins with a single species at time $0$ which forms a single lineage until a speciation event occurs at a random time $\tau$, where $\tau$ is exponentially distributed with rate parameter $\lambda$. At the speciation event the lineage splits into two lineages, both of which follow the same Yule process independently. For example, the phylogenetic $X$-tree in Figure~\ref{fig::simTree1} was generated by the sim.bd.taxa function in the Treesim R package with $n=10$, rate parameter $\lambda=0.416$ and death rate $\mu =0$. In particular, this call of function sim.bd.taxa generates a tree with $10$ taxa such that the sum of edge lengths from the root to any one taxon is constant. 
 We obtain the results in Table~\ref{tab::fig1} by using our algorithm for the RDDP from the previous section and solving all three problems in parallel, i.e., we evaluate ES and FP whenever we solve an instance of the MISP. Optimal solutions for variable $Y$ corresponding to the results in Table~\ref{tab::fig1} and measured by their respective phylogenetic diversity are shown in Table~\ref{tab::fig1.2}. Moreover, we find that our optimal solutions to the RDDP lie on the boundary of $\mathcal{P}(T,\ell)$, i.e., has coefficients $\gamma(x,e)\in\{0,1\}$ for all $x\in X$ and edges $e\in\{e_1,\dots,e_4\}$ in Figure~\ref{fig::simTree2}. Under objective function $\overline{\text{RDD}}_{\hat{T}}(\kappa)$ this is not the case. Specifically, the optimal solutions to the R3DP have coefficients $\gamma(x,e)\in\{0,1/4,1/2\}$ for all $x\in X$ and edges $e\in\{e_1,\dots,e_4\}$ in Figure~\ref{fig::simTree2}.

\begin{table*}[!t]
\centering
    \begin{tabular}{lcccc}
     \toprule
        $\kappa$ & $\text{PD}_{(T,\ell)}(Y_R^\kappa)/\kappa$ & $\text{PD}_{(T,\ell)}(Y_{DR}^{\kappa})/\kappa$ & $\text{PD}_{(T,\ell)}(Y_E^\kappa)/\kappa$ & $\text{PD}_{(T,\ell)}(Y_F^\kappa)/\kappa$\\ \midrule \\ [-1em]
        $2$ & $5.757$ & $\sim$ & $\sim$ & $\sim$ \\
        $3$ & $4.996$ & $\sim$ & $\sim$ & $\sim$ \\
        $4$ & $4.566$ & $\sim$ & $4.559$ & $4.584$ \\ \\[-1em]
        $5$ & $4.303$ & $4.322$ & $4.317$ & $4.322$ \\
        $6$ & $4.144$ & $\sim$ & $\sim$ & $\sim$\\ 
        $7$ & $3.717$ & $3.681$ & $3.611$ & $3.681$ \\ \\[-1em]
        $8$ & $2.942$ & $2.954$ & $3.247$ & $3.308$ \\ \\[-1em]
        $9$ & $2.987$ & $\sim$ & $3.015$ & $\sim$ \\
	\bottomrule \\[-1em]
    \end{tabular}
    \caption{The phylogenetic diversity (normalized by $\kappa$) of optimal solutions $Y_R^\kappa$, $Y_{DR}^{\kappa}$, $Y_E^\kappa$ and $Y_F^\kappa$ and  for the RDDP, R3DP, EDDP and the FDDP of order $\kappa$, respectively, for the phylogenetic $X$-tree $(T,\ell)$ in Figure~\ref{fig::simTree1}. The symbol $\sim$ indicates that the normalized phylogenetic diversity is equal to $\text{PD}_{(T,\ell)}(Y_R^\kappa)/\kappa$.}
    \label{tab::fig1.2}
\end{table*}
 
First, from Table~\ref{tab::fig1} we conclude for the phylogenetic $X$-tree $\hat{T}=(T,\ell)$ in Figure~\ref{fig::simTree1} that ES and FP do not satisfy bounds~\eqref{ineq::robustBound}. More specifically, diversity indices $\varphi_{\hat{T}}^{\text{ES}}$ and $\varphi_{\hat{T}}^{\text{FP}}$ are not robust with respect to neither uncertainty region $\mathcal{P}(T,\ell)$ nor $\mathcal{Q}(T,\ell)$. This means, when selecting a random set of $\kappa$ species to preserve, in the worst case an optimal solution to both the RDDP and R3DP captures higher levels of PD than ES or FP. Moreover, there is very little difference between optimal solutions $Y_R^\kappa$, $Y_E^\kappa$ and $Y_F^\kappa$ when measuring their effect on the phylogenetic diversity (Table~\ref{tab::fig1.2}). Hence, the advantage of a robust diversity index to capture higher levels of PD seems to be inherent to the index and to not depend on the detection of areas of the tree with high PD. Thus, overall our optimal solutions to the R3DP are the favoured indices because they provide a trade-off between being more robust than the ES and FP without being binary/too extreme. For example, for $\kappa=8$, our robust diversity index for uncertainty region $\mathcal{P}(T,\ell)$ is defined by coefficients $\gamma(x_7,e_1)=\gamma(x_{10},e_1)=1/2$ and $\gamma(x_4,e_2)=\gamma(x_8,e_3)=\gamma(x_3,e_4)=1$. Instead, our optimal solution to the R3DP modifies these latter binary coefficients, with $\gamma(x_4,e_2)=\gamma(x_2,e_3)=\gamma(x_3,e_4)=1/2$.


Secondly, we observe that the optimal solutions $Y_R^\kappa$ arise from selecting taxa which are spread out. That is, they form sets of taxa whose minimal connecting subtrees contain a large number of edges. For example, $Y_R^3=\{x_1,x_7,x_8\}$. This is not particularly surprising, as the diversity difference is smaller for sets whose diversity index scores draw from fewer edges (and is zero exactly when the $\kappa$ leaves form a monophyletic clade whose most recent common ancestor is adjacent to the root vertex). 
Another cause of the similarity between solutions $Y_R^\kappa$, $Y_{DR}^\kappa$, $Y_E^\kappa$ and $Y_F^\kappa$ could be simply that the dimension of $\mathcal{P}(T,\ell)$ is very low ($d_{\hat{T}}=2$). Hence, to better understand the differences between these diversity indices, we continue our analysis for larger trees. 

\begin{figure}[!t]
\centering
\includegraphics[scale=0.35]{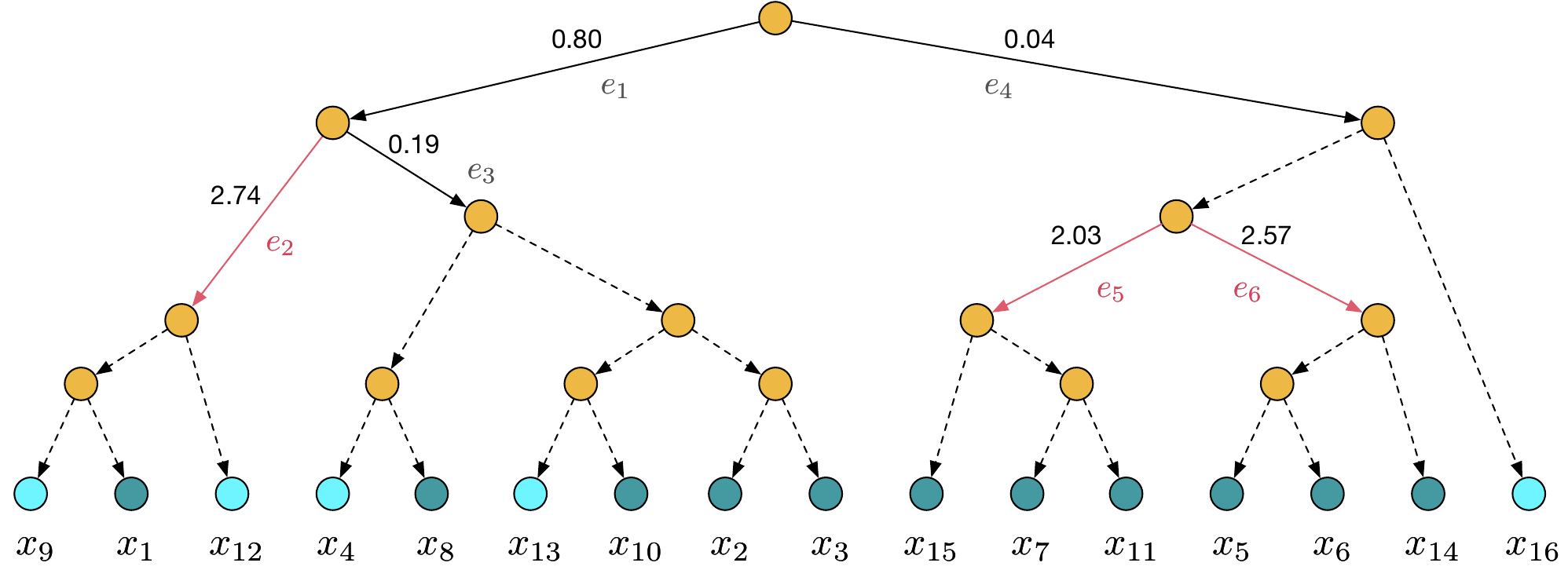}
\caption{A Yule tree $\hat{T}$ with $n=16$ taxa with most edge lengths suppressed for readability and dimension $d_{\hat{T}}=4$. Coefficients $\gamma(x,e)$, $x\in X$, are constant for consistent diversity indices for all dashed edges $e$. Red edges $\{e_2,e_5,e_6\}$ share a neutrality condition and light blue taxa form a set $Y_9=W_{12}\cup\{x_{12}\}$ (see Appendix~\ref{appendix:MISP}).}\label{fig::simTree16}
\end{figure}
\begin{table*}[!b]
\centering
    \begin{tabular}{llllllll}
     \toprule
        & \multicolumn{7}{l}{Species $\kappa$ selected in the Yule tree from Figure~\ref{fig::simTree16}}  \\ \cmidrule{2-8}
        & $2$ & $4$ & $6$ & $8$ & $10$ & $12$ & $14$ \\ \midrule
	$\text{RDD}_{\hat{T}}(\kappa)=$ & $3.780$ & $7.255$ & $8.081$ & $7.241$ & $5.656$ & $4.077$ & $1.639$ \\ \\ [-1em]
	$\overline{\text{RDD}}_{\hat{T}}(\kappa)=$ & $4.185$ & $7.531$ & $8.257$ & $7.507$ & $5.868$ & $4.086$ & $2.026$ \\ \\ [-1em]
	$\text{FDD}_{\hat{T}}(\kappa)=$ & $4.751$ & $7.939$ & $8.571$ & $7.852$ & $6.316$ & $4.451$ & $2.441$\\ \\[-1em]
	$\text{EDD}_{\hat{T}}(\kappa)=$ & $5.195$ & $8.613$ & $9.265$ & $8.579$ & $6.946$ & $5.104$ & $2.920$\\ 
	\midrule
	Time RDDP (sec.) & $0.018$ & $0.238$ & $0.969$ & $1.560$ & $0.880$ & $0.236$ & $0.015$ \\
	Time R3DP (sec.) & $0.025$ & $0.248$ & $1.165$ & $1.949$ & $1.049$ & $0.227$ & $0.015$ \\
	\bottomrule \\
    \end{tabular}
    \caption{The optimal objective function values for the RDDP, R3DP, FDDP and EDDP for different choices of~$\kappa$ for the phylogenetic $X$-tree $\hat{T}$ in Figure~\ref{fig::simTree16}. In addition, we note the solution time to solve the RDDP and R3DP of order $\kappa$ in seconds.}
    \label{tab::fig3}
\end{table*}

For our next instance, we use the phylogenetic $X$-tree in Figure~\ref{fig::simTree16}. This tree was generated by the same function and parameters as our tree in Figure~\ref{fig::simTree1} except for choosing $n=16$. Analogous results as in the last example are shown in Tables~\ref{tab::fig3} and~\ref{tab::fig3.2}. We can draw similar conclusions as for our previous case when $n=10$. Specifically, the optimal solutions to the RDDP and R3DP prove that neither ES nor FP are robust with respect to uncertainty regions $\mathcal{P}(T,\ell)$ or $\mathcal{Q}(T,\ell)$ (see Table~\ref{tab::fig3}). Furthermore, higher levels of PD capture do not (or to a very small degree) depend on the PD of particular subtrees (see Table~\ref{tab::fig3.2}). This means robust diversity indices for $\mathcal{Q}(T,\ell)$ are again favoured. For example, while the optimal solution to the RDDP of order~$5$ is almost fully binary, the coefficients for our optimal solution to the R3DP of order~$5$ take all values in $\{0,1/8,1/4,1/2\}$. Notice that this worst case analysis of the discussed indices does not reveal much about the ability of the indices to assess the shape of the tree. While ES and FP are based on criteria evaluating the tree shape directly, our robust diversity indices are not taking tree shape into account explicitly but seek to preserve very similar sets of taxa in the worst case. For example, for $\kappa=5$ we have 
\begin{align*}
&Y_E^5=Y_F^5=\{x_1,x_4,x_5,x_7,x_{10}\},~Y_R^5=\{x_1,x_4,x_7,x_{10},x_{14}\}\\
&\text{and }Y_{DR}^5=\{x_4,x_5,x_{10},x_{11},x_{12}\}.
\end{align*}

\begin{table*}[!t]
\centering
\begin{tabular}{lcccc}
     \toprule
        $\kappa$ & $\text{PD}_{(T,\ell)}(Y_R^\kappa)/\kappa$ & $\text{PD}(Y_{DR}^{\kappa})/\kappa$ & $\text{PD}(Y_E^\kappa)/\kappa$ & $\text{PD}(Y_F^\kappa)/\kappa$\\ \midrule \\ [-1em]
        $2$ & $3.582$ & $\sim$ & $\sim$ & $\sim$ \\ \\[-1em]
        $4$ & $3.309$ & $\sim$ & $\sim$ & $\sim$ \\ \\[-1em]
        $6$ & $2.932$ & $\sim$ & $\sim$ & $\sim$ \\ \\[-1em]
        $8$ & $2.766$ & $\sim$ & $\sim$ & $\sim$\\ \\[-1em]
        $10$ & $2.383$ & $2.412$ & $2.336$ & $2.460$\\ \\[-1em]
        $12$ & $2.113$ & $2.152$ & $2.006$ & $2.152$\\ \\[-1em]
        $14$ & $1.775$ & $1.849$ & $1.820$ & $1.849$\\
	\bottomrule \\[-1em]
    \end{tabular}
    \caption{The phylogenetic diversity (normalized by $\kappa$) of optimal solutions $Y_R^\kappa$, $Y_{DR}^{\kappa}$, $Y_E^\kappa$ and $Y_F^\kappa$ and  for the RDDP, R3DP, EDDP and the FDDP of order $\kappa$, respectively, for the phylogenetic $X$-tree $(T,\ell)$ in Figure~\ref{fig::simTree16}. The symbol $\sim$ indicates that the normalized phylogenetic diversity is equal to $\text{PD}_{(T,\ell)}(Y_R^\kappa)/\kappa$.}
    \label{tab::fig3.2}
\end{table*}

\begin{figure}[!b]
\centering
\includegraphics[scale=0.35]{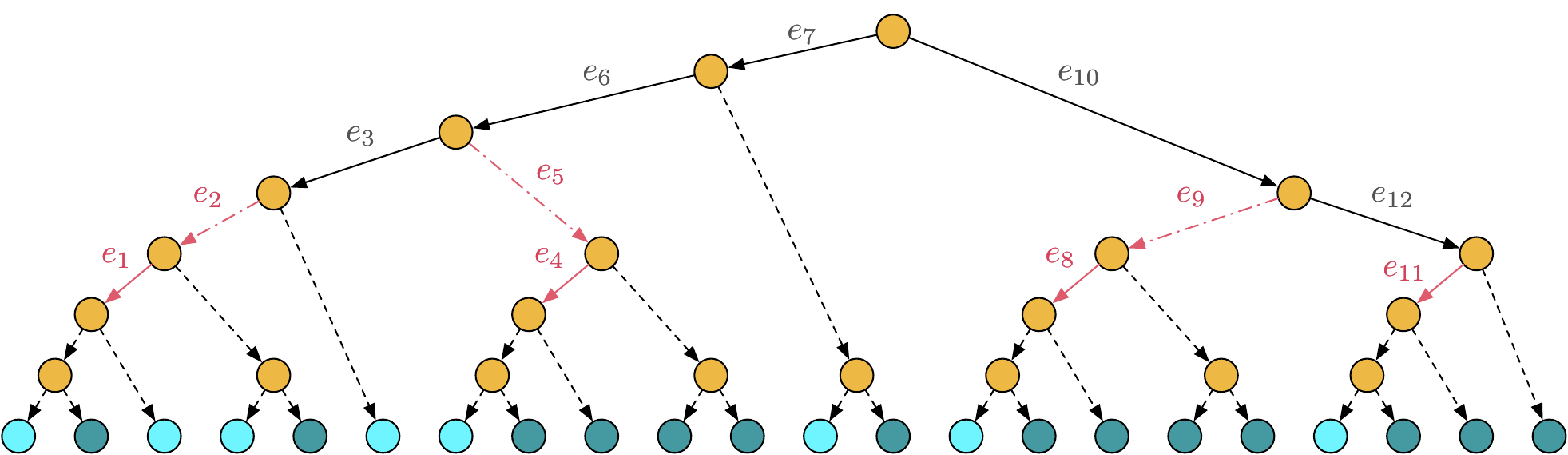}
\caption{A rooted phylogenetic $X$-tree $\hat{T}$ with $n=22$ taxa representing different species of albatross from birdtree.org. The taxa labels and edge lengths are suppressed for readability and we have dimension $d_{\hat{T}}=7$. Coefficients $\gamma(x,e)$, $x\in X$, are constant for consistent diversity indices for all dashed edges $e$. Both straight red edges $\{e_1,e_4,e_8,e_{11}\}$ and dashed red edges $\{e_2,e_5,e_9\}$ share a neutrality condition each and light blue taxa form a set which fits into Proposition~\ref{prop::FM}.}\label{fig::albaTree22}
\end{figure}

Last but not least, we apply our methods to a phylogenetic tree of albatrosses (family Diomedeidae).
The tree is derived from a set of data studied by Jetz et al.~\cite{jetz12} (see \url{http://www.birdtree.org}). The albatross family provides a good test case for the use of diversity indices for a number of reasons, not least being a high level of extinction risk. This is shown by the IUCN categorization of albatross species. Of the $22$ species contained in Figure~\ref{fig::albaTree22}, two are critically endangered, seven are endangered, six are vulnerable, six are near vulnerable, leaving only one rated of least concern~\cite{iucn06}. Additionally, while some protective measures can be implemented widely, such as those reducing incidental fisheries deaths, land-based albatross conservation requires interventions at the species level~\cite{Baker01032002,phillips2016conservation}. These birds often breed in remote island colonies, making their monitoring and management resource-intensive. A phylogenetic diversity index can therefore be an informative factor when prioritizing these resources.

\begin{table*}[!t]
\centering
    \begin{tabular}{llllllll}
     \toprule
        & \multicolumn{7}{l}{Species $\kappa$ selected in the tree from Figure~\ref{fig::albaTree22}}  \\ \cmidrule{2-8}
        & $2$ & $5$ & $8$ & $11$ & $14$ & $17$ & $20$ \\ \midrule
	$\text{RDD}_{\hat{T}}(\kappa)=$ & $26.789$ & $44.934$ & $47.592$ & $36.866$ & $31.356$ & $20.771$ & $11.634$ \\ \\ [-1em]
	$\overline{\text{RDD}}_{\hat{T}}(\kappa)=$ & $27.201$ & $44.934$ & $47.592$ & $37.258$ & $31.356$ & $20.771$ & $11.634$ \\ \\ [-1em]
	$\text{FDD}_{\hat{T}}(\kappa)=$ & $41.879$ & $59.498$ & $57.134$ & $49.024$ & $39.546$ & $29.292$ & $14.879$\\ \\ [-1em]
	$\text{EDD}_{\hat{T}}(\kappa)=$ & $45.123$ & $64.501$ & $61.650$ & $55.358$ & $45.751$ & $33.874$ & $16.246$\\ \midrule
	Time RDDP (sec.) & $0.065$ & $6.154$ & $61.613$ & $140.908$ & $63.742$ & $4.712$ & $0.041$\\
	Time R3DP (sec.) & $0.060$ & $5.295$ & $63.897$ & $151.637$ & $61.904$ & $5.062$ & $0.042$\\
	\bottomrule \\
    \end{tabular}
    \caption{The optimal objective function values for the RDDP, R3DP, FDDP and EDDP for different choices of~$\kappa$ for the phylogenetic $X$-tree in Figure~\ref{fig::albaTree22}. In addition, we note the solution time to solve the RDDP and R3DP of order $\kappa$ in seconds.}
    \label{tab::fig4}
\end{table*}
\begin{table*}[!b]
\centering
\resizebox{\textwidth}{!}{\begin{tabular}{lcccc}
     \toprule
        $\kappa$ & $\text{PD}_{(T,\ell)}(Y_R^\kappa)/\kappa$ & $\text{PD}_{(T,\ell)}(Y_{DR}^{\kappa})/\kappa$ & $\text{PD}_{(T,\ell)}(Y_E^\kappa0/\kappa$ & $\text{PD}_{(T,\ell)}(Y_F^\kappa)/\kappa$\\ \midrule \\ [-1em]
        $2$ & $26.018$ & $\sim$ & $\sim$ & $\sim$ \\ \\[-1em]
        $5$ & $19.113$ & $\sim$ & $\sim$ & $\sim$ \\ \\[-1em]
        $8$ & $13.715$ & $\sim$ & $\sim$ & $\sim$ \\ \\[-1em]
        $11$ & $10.608$ & $11.517$ & $10.296$ & $11.296$\\ \\[-1em]
        $14$ & $8.629$ & $\sim$ & $8.778$ & $9.559$\\ \\[-1em]
        $17$ & $8.027$ & $\sim$ & $7.581$ & $\sim$\\ \\[-1em]
        $20$ & $7.162$ & $\sim$ & $\sim$ & $\sim$ \\
	\bottomrule \\[-1em]
    \end{tabular}}
    \caption{The phylogenetic diversity (normalized by $\kappa$) of optimal solutions $Y_R^\kappa$, $Y_{DR}^{\kappa}$, $Y_E^\kappa$ and $Y_F^\kappa$ for the RDDP, R3DP, EDDP and FDDP of order $\kappa$, respectively, for the phylogenetic $X$-tree $(T,\ell)$ in Figure~\ref{fig::albaTree22}. The symbol $\sim$ indicates that the normalized phylogenetic diversity is equal to $\text{PD}_{(T,\ell)}(Y_R^\kappa)/\kappa$.}
    \label{tab::fig4.2}
\end{table*}

For this tree of real data we make similar observations as for the Yule trees on $10$ and $16$ taxa. In addition, notice that the optimal solution to the RDDP and the R3DP of order $\kappa$ differ only for $\kappa =2, 11$ (see Table~\ref{tab::fig4}). Even in those cases, they differ very slightly and preserve the same phylogenetic diversity (see Table~\ref{tab::fig4.2}). This lack of differentiation can be explained by the high number of balanced subtrees present in Figure~\ref{fig::albaTree22}. For example, 16 out of 22 taxa have a \emph{sibling}, i.e. are children of a shared common ancestor. This means, the imposition~\eqref{con::Peps} is redundant for most taxa. Furthermore, the increase in the dimension $d_{\hat{T}}$ from $4$ in Figure~\ref{fig::simTree16} to $7$ here does not seem to impact the propensity for robust diversity indices to be well spread out and very similar to the optimal solutions to the EDDP and FDDP. That said, we note that the sets obtained do not usually form a set that obtains the maximal PD score for sets of its size, as may be guessed. 

Overall, we expect the observed traits to persist for larger data sets. However, since the run time of our algorithm to solve the RDDP is a concave function in $\kappa$ and the increase in run time from Table~\ref{tab::fig3} to~\ref{tab::fig4} indicates poor scalability of this function, broader conclusions about the advantages and disadvantages of the optimal solution to the RDDP cannot be made at this point.

\subsection{Selecting diversity indices for uncertain tree shapes}
\begin{table*}[!t]
\centering
    \begin{tabular}{lllllllll}
     \toprule
        & \multicolumn{8}{l}{Species $\kappa$ selected in the tree from Figure~\ref{fig::simTree2}}  \\ \cmidrule{2-9}
        & $2$ & $3$ & $4$ & $5$ & $6$ & $7$ & $8$ & $9$ \\ \midrule
	$\theta_{\text{R}}=$ & $0.324$ & $0.298$ & $0.298$ & $0.324$ & $0.324$ & $0.324$ & $0.298$ & $0.298$ \\ \\ [-1em]
	$\theta_{\text{DR}}=$ & $0.324$ & $0.350$ & $0.350$ & $0.324$ & $0.324$ & $0.324$ & $0.350$ & $0.350$ \\ \midrule
	& \multicolumn{8}{l}{Species $\kappa$ selected in the tree from Figure~\ref{fig::simTree16}}  \\ \cmidrule{2-9}
        & $2$ & $4$ & $6$ & $8$ & $10$ & $12$ & $14$ & \\ \midrule
	$\theta_{\text{R}}=$ & $0.051$ & $0.051$ & $0.122$ & $0.122$ & $0.127$ & $0.122$ & $0.122$ & \\ \\ [-1em]
	$\theta_{\text{DR}}=$ & $0.692$ & $0.692$ & $0.738$ & $0.716$ & $0.741$ & $0.736$ & $0.736$ & \\ \midrule
	& \multicolumn{8}{l}{Species $\kappa$ selected in the tree from Figure~\ref{fig::albaTree22}}  \\ \cmidrule{2-9}
        & $2$ & $5$ & $8$ & $11$ & $14$ & $17$ & $20$ & \\ \midrule
	$\theta_{\text{R}}=$ & $0.221$ & $0.049$ & $0.185$ & $0.049$ & $0.049$ & $0.122$ & $0.049$ & \\ \\ [-1em]
	$\theta_{\text{DR}}=$ & $0.338$ & $0.193$ & $0.328$ & $0.193$ & $0.328$ & $0.266$ & $0.193$ & \\
	\bottomrule \\
    \end{tabular}
    \caption{The maximum choice of $\theta$ for which the optimal solution to the RDDP and R3DP is $\theta$-compatible denoted by $\theta_R$ and $\theta_{\text{DR}}$, respectively.}
    \label{tab::theta1}
\end{table*}
We continue our analysis of the optimal solutions to robust biodiversity problems by considering the rooted phylogenetic $X$-trees in Figures~\ref{fig::simTree1},~\ref{fig::simTree16} and~\ref{fig::albaTree22} as inputs for the MCDP. Throughout the experiments we choose $\epsilon =10^{-7}$ to define the objective function of the MCDP.

First, we consider the $\theta$-compatibility of the diversity indices discussed in the last subsection. Table~\ref{tab::theta1} shows this information for the RDDP and R3DP. We know that FP is always $1$-compatible and we can calculate that ES is $\theta_{\text{ES}}$-compatible for a maximum choice for $\theta_{\text{ES}}$ of $0.534$, $0.755$ and $0.714$ for the tree in Figure~\ref{fig::simTree1},~\ref{fig::simTree16} and~\ref{fig::albaTree22}, respectively. Since high (low) values of $\theta$ indicate that the distances $|\gamma(x_i,e)-\gamma(x_j,e)|$ between non-zero coefficients $\gamma(x_i,e),\gamma(x_i,e),\,e\in E,\,x_i,x_j\in X$ are small (large), we see that ES is more continuous/more closely related to FP than both the optimal solution to the RDDP and R3DP. Unsurprisingly, the R3DP exhibits higher levels of compatibility than the RDDP because conditions~\eqref{con::Peps} directly limit the lower bounds~\eqref{comp::con1}. However, only for our second tree (see Figure~\ref{fig::simTree16}) is this gap substantial. A potential explanation for this discrepancy between the three trees is the distinctiveness in tree shape. Since the second tree contains a larger proportion of pairwise distinct tree shapes among its subtrees than the other two trees, bounds~\eqref{comp::con1} are less uniform for the second tree, yielding a larger left-hand side in constraint~\eqref{comp::con3}.\\

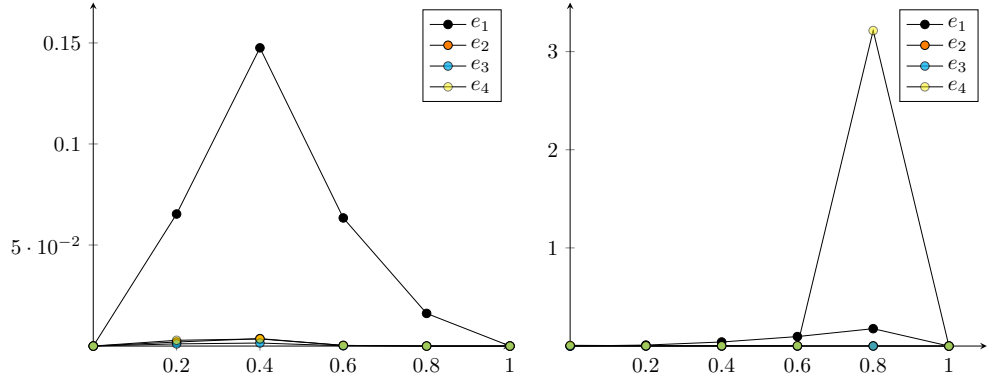
\begin{figure}[!t]
\centering
\begin{tikzpicture}[scale=0.8]
\begin{axis}[
    axis lines=middle,
    xmin=0, xmax=1,
    ymin=0, ymax=0.17,
]
\addplot [mark=*, mark options={fill=black}] table {
1 0
0.8 0.0161470671
0.6 0.0633984687
0.4 0.1475660742
0.2 0.0653395013
0 0
};
\addplot [mark=*, mark options={fill=orange,opacity=1}] table {
1 0
0.8 0
0.6 0.00033191
0.4 0.003673144
0.2 0.001992148
0 0
};
\addplot [mark=*, mark options={fill=cyan,opacity=0.75}] table {
1 0
0.8 0
0.6 0.000127079
0.4 0.001495075
0.2 0.001075835
0 0
};
\addplot [mark=*, mark options={fill=yellow,opacity=0.5}]  table {
1 0
0.8 0
0.6 0.000317291
0.4 0.003523098
0.2 0.002846115
0 0
};
\legend{$e_1$,$e_2$,$e_3$, $e_4$};
\end{axis}
\end{tikzpicture}~
\begin{tikzpicture}[scale=0.8]
\begin{axis}[
    axis lines=middle,
    xmin=0, xmax=1.1,
    ymin=0, ymax=3.5,
]
\addplot [mark=*, mark options={fill=black}] table {
1 0
0.8 0.177017159
0.6 0.096578428
0.4 0.041290604
0.2 0.009953365
0 0
};
\addplot [mark=*, mark options={fill=orange,opacity=1}] table {
1 0
0.8 0.000928046
0.6 0
0.4 0
0.2 0
0 0.001269084
};
\addplot [mark=*, mark options={fill=cyan,opacity=0.75}] table {
1 0
0.8 0.00049414
0.6 0.001269084
0.4 0.001269084
0.2 0.001269084
0 0
};
\addplot [mark=*, mark options={fill=yellow,opacity=0.5}]  table {
1 0
0.8 3.214274298
0.6 0.003659806
0.4 0.003659806
0.2 0.003659806
0 0.00741791
};
\legend{$e_1$,$e_2$,$e_3$, $e_4$};
\end{axis}
\end{tikzpicture}
\caption{On the left the degree of discontinuity $\text{dd}(T,\ell,3,\theta,e)$ (rescaled by factor $10^{-1}$ for $e=e_1$) for the phylogenetic $X$-tree $(T,\ell)$ in Figure~\ref{fig::simTree2}, edges $e\in\{e_1,e_2,e_3,e_4\}$ and parameter $\theta$ on the $x$-axis. The graph on the right shows $\text{dd}(T,\ell,7,\theta,e)$ for the same set of parameters.}\label{fig::expComp1}
\end{figure}

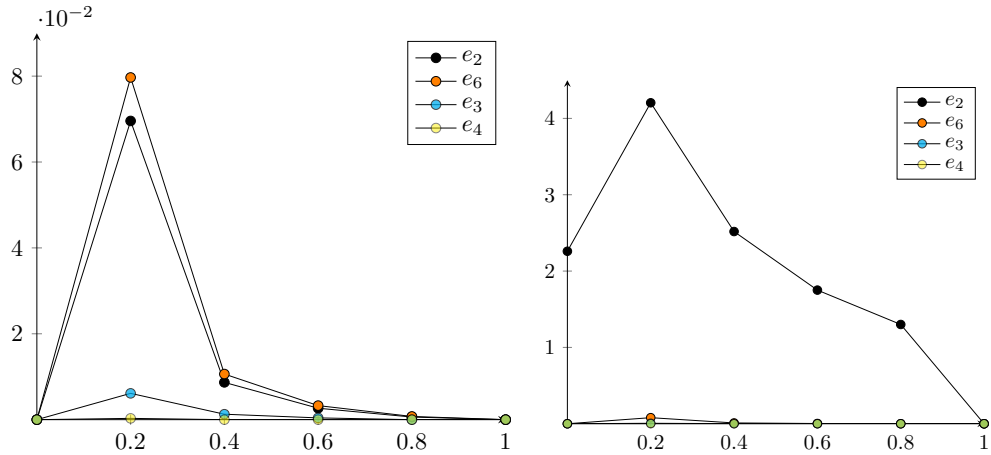
\begin{figure}[!b]
\centering
\begin{tikzpicture}[scale=0.9]
\begin{axis}[
    axis lines=middle,
    xmin=0, xmax=1,
    ymin=0, ymax=0.09,
]
\addplot [mark=*, mark options={fill=black}] table {
1 0
0.8 0.000609104
0.6 0.002665058
0.4 0.008682244
0.2 0.06956446
0 0
};
\addplot [mark=*, mark options={fill=orange,opacity=1}] table {
1 0
0.8 0.000749909
0.6 0.003276274
0.4 0.010624605
0.2 0.079691378
0 0
};
\addplot [mark=*, mark options={fill=cyan,opacity=0.75}] table {
1 0
0.8 0
0.6 0.000412745
0.4 0.001269119
0.2 0.006127082
0 0
};
\addplot [mark=*, mark options={fill=yellow,opacity=0.5}]  table {
1 0
0.8 0
0.6 0
0.4 0
0.2 0.000301125
0 0
};
\legend{$e_2$,$e_6$,$e_3$, $e_4$};
\end{axis}
\end{tikzpicture}~
\begin{tikzpicture}[scale=0.8]
\begin{axis}[
    axis lines=middle,
    xmin=0, xmax=1,
    ymin=0, ymax=4.5,
]
\addplot [mark=*, mark options={fill=black}] table {
1 0
0.8 1.300407021
0.6 1.750720646
0.4 2.51787187
0.2 4.204429636
0 2.259796086
};
\addplot [mark=*, mark options={fill=orange,opacity=1}] table {
1 0
0.8 0.000504932
0.6 0.003276274
0.4 0.010624605
0.2 0.079691378
0 0
};
\addplot [mark=*, mark options={fill=cyan,opacity=0.75}] table {
1 0
0.8 0.000236691
0.6 0.000412745
0.4 0.001269119
0.2 0.006127082
0 0
};
\addplot [mark=*, mark options={fill=yellow,opacity=0.5}]  table {
1 0
0.8 0
0.6 0
0.4 0
0.2 0.000301125
0 0
};
\legend{$e_2$,$e_6$,$e_3$, $e_4$};
\end{axis}
\end{tikzpicture}
\caption{On the left the degree of discontinuity $\text{dd}(T,\ell,4,\theta,e)$ for the phylogenetic $X$-tree $(T,\ell)$ in Figure~\ref{fig::simTree16}, edges $e\in\{e_2,e_6,e_3,e_4\}$ and parameter $\theta$ on the $x$-axis. The graph on the right shows $\text{dd}(T,\ell,12,\theta,e)$ for the same set of parameters.}\label{fig::expComp2}
\end{figure}

\begin{figure}[!t]
\centering
\begin{tikzpicture}[scale=0.9]
\begin{axis}[
    axis lines=middle,
    xmin=0, xmax=1.1,
    ymin=0, ymax=4,
]
\addplot [mark=*, mark options={fill=black}] table {
1 0
0.8 2.546819328
0.6 2.946933657
0.4 3.172238906
0.2 3.446933657
0 3.446933657
};
\addplot [mark=*, mark options={fill=orange,opacity=1}] table {
1 0
0.8 3.367357915
0.6 1.869064398
0.4 0.25
0.2 0
0 0
};
\addplot [mark=*, mark options={fill=cyan,opacity=0.75}] table {
1 0
0.8 0.092512121
0.6 0.395965659
0.4 0.509404881
0.2 1.034099632
0 1.034099632
};
\addplot [mark=*, mark options={fill=yellow,opacity=0.5}]  table {
1 0
0.8 1.364243233
0.6 0.111866026
0.4 0.225305248
0.2 0.25
0 0.25
};
\legend{$e_6$,$e_{10}$,$e_2$, $e_{11}$};
\end{axis}
\end{tikzpicture}~
\begin{tikzpicture}[scale=0.8]
\begin{axis}[
    axis lines=middle,
    xmin=0, xmax=1.1,
    ymin=0, ymax=3.7,
]
\addplot [mark=*, mark options={fill=black}] table {
1 0
0.8 0
0.6 0
0.4 0
0.2 0
0 0
};
\addplot [mark=*, mark options={fill=orange,opacity=1}] table {
1 0
0.8 0.404493384
0.6 0
0.4 0
0.2 0
0 0
};
\addplot [mark=*, mark options={fill=cyan,opacity=0.75}] table {
1 0
0.8 3.295386586
0.6 0
0.4 0
0.2 0
0 0
};
\addplot [mark=*, mark options={fill=yellow,opacity=0.5}]  table {
1 0
0.8 2.561323203
0.6 0.571826525
0.4 0
0.2 0
0 0
};
\legend{$e_6$,$e_{10}$,$e_2$, $e_{11}$};
\end{axis}
\end{tikzpicture}
\caption{On the left the degree of discontinuity $\text{dd}(T,\ell,5,\theta,e)$ for the phylogenetic $X$-tree $(T,\ell)$ in Figure~\ref{fig::albaTree22}, edges $e\in\{e_6,e_{10},e_2,e_{11}\}$ and parameter $\theta$ on the $x$-axis. The corresponding edge lengths are the two largest ($\ell(e_6)=14.613$, $\ell(e_{10})=10.755$) and two smallest ($\ell(e_2)=0.934$, $\ell(e_{11})=0.694$) among edges $\{e_1,\dots,e_{12}\}$. The graph on the right shows $\text{dd}(T,\ell,17,\theta,e)$ for the same set of parameters.}\label{fig::expComp3}
\end{figure}
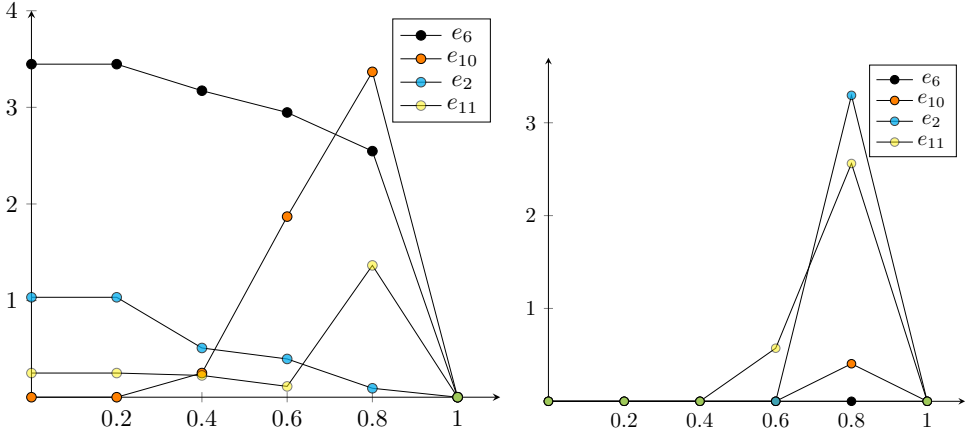

\noindent On the converse, we can fix both parameters $\kappa$ and $\theta$ and consider the MCDP of order $(\kappa,\theta)$. This approach yields a family of diversity indices which we want to characterize by the degree of discontinuity $\text{dd}(T,\ell,\kappa,\theta,e)$ (see Section~\ref{sec:1}). Figures~\ref{fig::expComp1},~\ref{fig::expComp2} and~\ref{fig::expComp3} show the degree of discontinuity for the phylogenetic $X$-trees in Figures~\ref{fig::simTree1},~\ref{fig::simTree16} and~\ref{fig::albaTree22}, respectively, and some fixed choices for parameters $\kappa$ and $\theta$.

First, observe that the degree of discontinuity attains its maximum for most edges under consideration for the same value of $\theta$. For example $\theta=0.4$ on the left in Figure~\ref{fig::expComp1} and $\theta=0.8$ on the right in the same figure. Since the respective edges are not concentrated in one particular subtree of $T$, we deduce that the change in the optimal solution to the MCDP when contracting one edge does depend on the structure of space $\mathcal{P}(T,\ell,\theta)$ but does not tend to rely on the tree shape of $T$. Clearly, this is not the case for $\text{dd}(T,\ell,5,\theta,e)$ in Figure~\ref{fig::expComp3} but a prevalent trend in all other of our experiments. Moreover, most of our experiments showed that sensitivity of a $\theta$-compatible diversity index maximizing ID to the collapse of an edge in the underlying tree is concave in $\theta$. This means, as $\theta$ decreases an optimal solution to the MCDP becomes more sensitive towards uncertainties in the tree shape until a threshold value of $\theta$. If this threshold value is larger than zero, then decreasing $\theta$ further removes restrictions on the set of optimal solutions to the MCDP which were imposed by larger values for $\theta$ / stricter requirements for relaxed continuity. Hence, the observation that $\text{dd}(T,\ell,\kappa,\theta,e)=0$ for small values of $\theta$ does not mean that looser relaxations of continuity are less sensitive to an uncertain tree shape than stricter requirements for relaxed continuity. Instead, we have to take into account that the degree of discontinuity measures the sensitivity of a $\theta$-compatible diversity index with respect to ID. This means, for changes of edge lengths which shift maxima of ID but do not rescale the underlying diversity index coefficients in the absence of tight $\theta$-compatibility constraints, the degree of discontinuity stays at zero.

Secondly, notice that the four edges in each figure consist of both two edges with the largest and smallest edge lengths, respectively, among all edges $e$ with non-constant non-zero coefficients $\gamma(x,e)$. Figures~\ref{fig::expComp1} and~\ref{fig::expComp2} indicate a correlation between the magnitude of the degree of discontinuity and the size of $\ell(e)$. However, this hypothesis is not supported for our tree on $22$ taxa (see the graph on the right in Figure~\ref{fig::expComp3}) because, for $\theta=0.8$, the optimal solution to the MCDP is much more sensitive to the contraction of the two shortest edges than the two longest edges. However, the underlying diversity indices reveal that for both edges $e_2$ and $e_{11}$ the increase in the degree of discontinuity solely relies on changes in coefficients $\gamma(x,e_6)$ for $e_6=(u,v)$, $x\in X[v]$. Thus, all experiments support that higher levels of discontinuity are caused by large edge lengths either directly or indirectly.

\section{Discussion and conclusions}\label{secRemarks}
In practice, current usage of diversity indices is restricted to the Fair Proportion (FP) and Equal-Splits (ES) indices. These measures were built from some fairly natural, self-contained assumptions regarding how evolutionary isolation ought to be measured on phylogenetic trees. The `naturalness' of these diversity indices has contributed to their dominance, particularly so in the case of FP. However, natural assumptions do not necessarily guarantee the best measures. This is indicated in part by the differences that do exist between FP and ES. Wicke and Steel \cite{wicke20} showed that these two diversity indices give the same values only for a class of very balanced phylogenetic trees. Additionally, that paper detailed quite large differences between these two measures in some cases. It can even happen that the diversity index scores of these two approaches differ enough to rank species in completely opposite orders \cite{manson24}. 

The description of the convex space $\mathcal{P}(T,\ell)$ associated with each phylogenetic tree allows us to find phylogenetic diversity indices beyond those that we can invent by following natural assumptions. 
In this work we outlined a method for finding a robust diversity index which captures more PD than either FP or ES in the context of a worst case analysis. The new robust indices can be used when it is desirable that the diversity of the individuals closely matches the diversity of the sets that they may form. 
However, our method does not fully reveal which characteristics of robust diversity indices are beneficial in practice/on average and future work comparing robust diversity indices across trees would improve our understanding of the diversity difference problem.

In addition, we introduced a subspace $\mathcal{Q}(T,\ell)\subset \mathcal{P}(T,\ell)$ of diversity index scores to illustrate how information about the tree shape (see~\eqref{con::Peps}) can enhance the usefulness of robust diversity indices. Therefore, adapting such subspaces of diversity indices to prevalent features of a class of phylogenetic $X$-trees offers new selection mechanisms for robust diversity indices which could be well-suited for conservation decision problems in practice.

Furthermore, we have shown that the notion of $\theta$-compatibility relaxes the continuity property of diversity indices and yields a hierarchy of distinct subspaces $\mathcal{P}(T,\ell,\theta)\subset\mathcal{P}(T,\ell)$. While we compared these subspaces using the degree of discontinuity, a measure biased by sets of maximum ID, we observed that large edge lengths drive changes in the diversity index selection under an uncertain tree shape. Hence, combining the notion of $\theta$-compatibility with assumptions on the features of long edges in a phylogenetic $X$-tree offers a control mechanism to maintain a scoring of taxa in the presence of uncertain edge lengths.

Finally, the different properties of diversity indices characterized by our introduction of certain subspaces could be used in conjunction, i.e., one could seek to find a robust diversity index in $\mathcal{Q}(T,\ell)\cap\mathcal{P}(T,\ell,\theta)$. This approach would allow for a more nuanced analysis but also requires broader large scale experiments. At this point the latter necessitates further improvements of the algorithms we have employed (see Section~\ref{sec3}).

In total, we have highlighted multiple directions of a properties-first strategy that may be further adapted to optimize for a number of diversity index features and conditions. We hope that this approach helps to continue the discussions about which properties are useful for diversity indices to possess or mitigate against. The optimization methods give a means of examining all of the possible indices for a best fit, unearthing methods that would almost certainly never come to light if only following `natural' assumptions.

\section*{Acknowledgments}
The first author acknowledges support from grant OCENW.M.21.306 from the Dutch Research Council (NWO). The second author acknowledges support from the New Zealand Marsden Fund (MFP-UOC2005).

\begin{appendices}

\section{A polynomial time exact solution algorithm for the MISP}\label{appendix:MISP}

In this section we describe a compact polynomial-sized LP formulation for the MISP to solve it by linear programming. For a rooted phylogenetic $X$-tree $\hat{T}=(T,\ell)$ and $Y\subseteq X$, recall that the MISP of $Y$ is defined by
\begin{align*}
\text{MIS}_{\hat{T}}(Y)=\min_{s\in\mathcal{P}(T,\ell)}\Delta_{\hat{T}}(Y,s).
\end{align*}
We start by modifying the input instance~$Y$: if there exist taxa $y,z\in X[v]$ for some $v\in V(T)$ such that $T[v]$ is balanced, then remove~$z$ from~$Y$ and repeat until there is no more valid choice for $v$. For the updated set $Y'$, let $\omega(Y')\in\mathbb{R}^{Y'}$ be defined for each entry~$i$ as one plus the number of times taxon $x_i\in Y'$ was not removed from a pair of taxa checked in our removal procedure for~$Y$. 

Next, we know that there exist $d_{(T,\ell)}$ diversity index scores in $s\in\mathcal{P}(T,\ell)$ which fully determine $s$ (see Proposition~\ref{prop::FM}). Let $S(T,\ell)\subset X^{d_{(T,\ell)}}$ denote the set of sets of $d_{(T,\ell)}$ taxa corresponding to these diversity index scores and let $\mathcal{Z}(T,\ell)$ denote the set of subsets $Z$ fitting into~\eqref{facet2}, i.e., $S(T,\ell)\subset\mathcal{Z}(T,\ell)$. 
Clearly, for $W\in S(T,\ell)$, calculating diversity index scores of taxa in $W$ does fully determine $\sum_{x_i\in Y'}s_i$ for any $s\in\mathcal{P}(T,\ell)$. In other words, there exists a linear function $f_{Y'\setminus W}:\mathbb{R}^{W}\to\mathbb{R}$ such that
\begin{align}\label{eq::Y'W}
\sum_{x_i\in Y'}s_i=f_{Y'\setminus W}\left(\left(s_i-\,\text{LB}(x_i,T)\right)_{x_i\in W}\right)+\sum_{x_i\in W}s_i.
\end{align}
Here, we shift diversity index scores $s_i$ by $\,\text{LB}(x_i,T)$ to simplify our notation later in this section. Identity~\eqref{eq::Y'W} will be useful when solving the MISP. Further helpful definitions are the following:
let $W_k\in S(T,\ell)$ denote the sets of taxa for which Proposition~\ref{prop::FM} applies to $Y_i=W_k\cup\{x_k\}$ and let $\mathcal{S}\left(\hat{T},Y_i\setminus\{x_k\}\right)$ denote the set of subsets of $W_k$ in $\mathcal{Z}(T,\ell)$ with maximum cardinality among all possible choices for $k$. We illustrate $\mathcal{S}\left(\hat{T},Y_i\setminus\{x_k\}\right)$ in the following example:
\begin{example}\label{ex1}
Consider $W_4=\{x_2,x_5\}$ in Figure~\ref{fig::simTree2}. Then, $Y_i=W_4\cup\{x_4\}$ for $i\in\{2,5\}$. Furthermore, 
\begin{align*}
\mathcal{Z}(x_4,Y_5)=\left\{\{x_2,x_5\}\right\}~~~\text{and}~~~\mathcal{Z}(x_4,Y_2)=\left\{\{x_2\},\{x_2,x_5\}\right\}.
\end{align*}
Notice that $\{x_2\},\{x_5\}\notin\mathcal{Z}(x_4,Y_5)$ because $x_5\notin\{x_2\}$ and $F(Y_5-\{x_5\})$ is not a tree, respectively. Similarly, $\{x_5\}\notin\mathcal{Z}(x_4,Y_2)$ because $x_2\notin\{x_5\}$. Next, consider $W_8=\{x_2,x_5\}$ in the same figure. Then, $Y_i=W_8\cup\{x_8\}$ for $i\in\{2,5\}$ and we get $\mathcal{Z}(x_8,Y_2)=\left\{\{x_2,x_5\}\right\}$. The same argument can be repeated for $W_3=\{x_2,x_5\}$. Thus,
\begin{align*}
\mathcal{S}\left(\hat{T},\{x_2,x_5\}\right)=\left\{\{x_2\},\{x_2,x_5\}\right\}.
\end{align*}
Moreover, for $Y'=\{x_2,x_4,x_5\}$, $W=\{x_2,x_5\}$ and $s\in\mathcal{P}(T,\ell)$, we have
\begin{align*}
&f_{Y'\setminus W}\left(\left(s_i-\,\text{LB}(x_i,T)\right)_{x_i\in W}\right)\\
&=f_{\{x_4\}}\left(\left(s_i-\,\text{LB}(x_i,T)\right)_{x_i\in W}\right)\\
&=\,\text{LB}(x_4,T)+0.2\cdot\left(1-\frac{s_2-\,\text{LB}(x_2,T)}{0.13/2+2.28/4}\right)+4.85-\left(s_5-\,\text{LB}(x_5,T)\right).
\end{align*}
\end{example}
Finally, by substituting variables $s\in\mathcal{P}(T,\ell)$ entrywise with variables $t_i=s_i-\,\text{LB}(x_i,T)$ and given $Y\subseteq X$, $W\in S(T,\ell)$, we arrive at:
\begin{form}\label{form::MISP}
\begin{align*}
\text{MIS}_{\hat{T}}(Y)=\,\text{PD}_{\hat{T}}(Y)-\sum_{x_j\in Y}\text{LB}(x_j,T)~~~~~~~~~~~~&\\
-\,\max~f_{Y'\setminus W}(t)+\sum_{x_j\in W}\omega(W)_j\cdot t_j&\\
\text{s.t.}\,~~~~~~~~~~~~~~~~~~~~\sum_{x_j\in Z}N_j\cdot t_j&\leq r_Z~&~&\forall\,Z\in\mathcal{S}\left(\hat{T},W\right)\\
t&\in\mathbb{R}_{\geq 0}^W
\end{align*}
\end{form}
In general, we conclude that, for any $W\in\mathcal{S}(T,\ell)$, the size of set $\mathcal{S}\left(\hat{T},W\right)$ grows at most linearly in $|W|$. Specifically, if the sets of taxa in $\mathcal{Z}(x_k,Y)$ induce a \emph{caterpillar}, i.e., the internal vertices form one directed path, then $|\mathcal{Z}(x_k,Y)|=|W_k|$. Thus, Problem~\ref{prob::MISP} can be solved in polynomial time with significantly less constraints than in a compact description of $\mathcal{P}(T,\ell)$. This can be done by hand for sufficiently small $d_{(T,\ell)}$ as the following example illustrates. Otherwise, any LP solver can produce an optimal solution.
\begin{example}\label{ex1::2}
Continuing Example~\ref{ex1}, for $\hat{T}$ in Figure~\ref{fig::simTree2} we have
\begin{align*}
f_{\{x_4\}}\left(t_2,t_5\right)&=\frac{101}{20}-\frac{40}{127}t_2-t_5.
\end{align*}
Hence, for $W=\{x_2,x_5\}$ we deduce from Formulation~\ref{form::MISP} that
\begin{align*}
&\text{MIS}_{\hat{T}}\left(\{x_2,x_4,x_5\}\right)\\
&=12.41-4.85-\max\left\{f_{\{x_4\}}(t_2,t_5)+t_2+t_5\,:\,t_2\in[0,0.635],~t_2+t_5\in[0,5.05]\right\}\\
&=12.41-4.85-(0+0.635+4.85)=2.075.
\end{align*}
\end{example}

\end{appendices}


\begin{thebibliography}{32}
\ifx \bisbn   \undefined \def \bisbn  #1{ISBN #1}\fi
\ifx \binits  \undefined \def \binits#1{#1}\fi
\ifx \bauthor  \undefined \def \bauthor#1{#1}\fi
\ifx \batitle  \undefined \def \batitle#1{#1}\fi
\ifx \bjtitle  \undefined \def \bjtitle#1{#1}\fi
\ifx \bvolume  \undefined \def \bvolume#1{\textbf{#1}}\fi
\ifx \byear  \undefined \def \byear#1{#1}\fi
\ifx \bissue  \undefined \def \bissue#1{#1}\fi
\ifx \bfpage  \undefined \def \bfpage#1{#1}\fi
\ifx \blpage  \undefined \def \blpage #1{#1}\fi
\ifx \burl  \undefined \def \burl#1{\textsf{#1}}\fi
\ifx \doiurl  \undefined \def \doiurl#1{\url{https://doi.org/#1}}\fi
\ifx \betal  \undefined \def \betal{\textit{et al.}}\fi
\ifx \binstitute  \undefined \def \binstitute#1{#1}\fi
\ifx \binstitutionaled  \undefined \def \binstitutionaled#1{#1}\fi
\ifx \bctitle  \undefined \def \bctitle#1{#1}\fi
\ifx \beditor  \undefined \def \beditor#1{#1}\fi
\ifx \bpublisher  \undefined \def \bpublisher#1{#1}\fi
\ifx \bbtitle  \undefined \def \bbtitle#1{#1}\fi
\ifx \bedition  \undefined \def \bedition#1{#1}\fi
\ifx \bseriesno  \undefined \def \bseriesno#1{#1}\fi
\ifx \blocation  \undefined \def \blocation#1{#1}\fi
\ifx \bsertitle  \undefined \def \bsertitle#1{#1}\fi
\ifx \bsnm \undefined \def \bsnm#1{#1}\fi
\ifx \bsuffix \undefined \def \bsuffix#1{#1}\fi
\ifx \bparticle \undefined \def \bparticle#1{#1}\fi
\ifx \barticle \undefined \def \barticle#1{#1}\fi
\bibcommenthead
\ifx \bconfdate \undefined \def \bconfdate #1{#1}\fi
\ifx \botherref \undefined \def \botherref #1{#1}\fi
\ifx \url \undefined \def \url#1{\textsf{#1}}\fi
\ifx \bchapter \undefined \def \bchapter#1{#1}\fi
\ifx \bbook \undefined \def \bbook#1{#1}\fi
\ifx \bcomment \undefined \def \bcomment#1{#1}\fi
\ifx \oauthor \undefined \def \oauthor#1{#1}\fi
\ifx \citeauthoryear \undefined \def \citeauthoryear#1{#1}\fi
\ifx \endbibitem  \undefined \def \endbibitem {}\fi
\ifx \bconflocation  \undefined \def \bconflocation#1{#1}\fi
\ifx \arxivurl  \undefined \def \arxivurl#1{\textsf{#1}}\fi
\csname PreBibitemsHook\endcsname

\bibitem[\protect\citeauthoryear{Myers}{1989}]{myers89}
\begin{bchapter}
\bauthor{\bsnm{Myers}, \binits{N.}}:
\bctitle{A major extinction spasm: predictable and inevitable?}
In: \beditor{\bsnm{Western}, \binits{D.}},
\beditor{\bsnm{Pearl}, \binits{M.C.}} (eds.)
\bbtitle{Conservation for the Twenty-first Century},
p. \bfpage{49}.
\bpublisher{Oxford University Press},
\blocation{New York}
(\byear{1989})
\end{bchapter}
\endbibitem

\bibitem[\protect\citeauthoryear{Singh}{2002}]{Singh02}
\begin{barticle}
\bauthor{\bsnm{Singh}, \binits{J.S.}}:
\batitle{The biodiversity crisis: a multifaceted review}.
\bjtitle{Current Science}
\bvolume{82}(\bissue{6}),
\bfpage{638}--\blpage{647}
(\byear{2002})
\end{barticle}
\endbibitem

\bibitem[\protect\citeauthoryear{Gaston and Spicer}{2013}]{spicer13}
\begin{bbook}
\bauthor{\bsnm{Gaston}, \binits{K.J.}},
\bauthor{\bsnm{Spicer}, \binits{J.I.}}:
\bbtitle{Biodiversity: {A}n Introduction}.
\bpublisher{John Wiley \& Sons},
\blocation{Oxford}
(\byear{2013})
\end{bbook}
\endbibitem

\bibitem[\protect\citeauthoryear{IUCN}{2006}]{iucn06}
\begin{botherref}
\oauthor{\bsnm{IUCN}}:
Red list of threatened species.
http://www.iucnredlist.org
(2006)
\end{botherref}
\endbibitem

\bibitem[\protect\citeauthoryear{Vane-Wright et~al.}{1991}]{wright90}
\begin{barticle}
\bauthor{\bsnm{Vane-Wright}, \binits{R.I.}},
\bauthor{\bsnm{Humphries}, \binits{C.J.}},
\bauthor{\bsnm{Williams}, \binits{P.H.}}:
\batitle{What to protect? systematics and the agony of choice}.
\bjtitle{Biological Conservation}
\bvolume{55},
\bfpage{235}--\blpage{254}
(\byear{1991})
\end{barticle}
\endbibitem

\bibitem[\protect\citeauthoryear{Crozier}{1997}]{crozier97}
\begin{barticle}
\bauthor{\bsnm{Crozier}, \binits{R.H.}}:
\batitle{Preserving the information content of species: genetic diversity,
  phylogeny, and conservation worth}.
\bjtitle{Annual Review of Ecology and Systematics}
\bvolume{28},
\bfpage{243}--\blpage{268}
(\byear{1997})
\end{barticle}
\endbibitem

\bibitem[\protect\citeauthoryear{Hartmann and Steel}{2006}]{hartmann06}
\begin{barticle}
\bauthor{\bsnm{Hartmann}, \binits{K.}},
\bauthor{\bsnm{Steel}, \binits{M.}}:
\batitle{Maximizing phylogenetic diversity in biodiversity conservation: greedy
  solutions to the noah's ark problem}.
\bjtitle{Systematic Biology}
\bvolume{55}(\bissue{4}),
\bfpage{644}--\blpage{651}
(\byear{2006})
\end{barticle}
\endbibitem

\bibitem[\protect\citeauthoryear{Tucker et~al.}{2017}]{tucker17}
\begin{barticle}
\bauthor{\bsnm{Tucker}, \binits{C.M.}},
\bauthor{\bsnm{Cadotte}, \binits{M.W.}},
\bauthor{\bsnm{Carvalho}, \binits{S.B.}},
\bauthor{\bsnm{Davies}, \binits{T.J.}},
\bauthor{\bsnm{Ferrier}, \binits{S.}},
\bauthor{\bsnm{Fritz}, \binits{S.A.}},
\bauthor{\bsnm{Grenyer}, \binits{R.}},
\bauthor{\bsnm{Helmus}, \binits{M.R.}},
\bauthor{\bsnm{Jin}, \binits{L.S.}},
\bauthor{\bsnm{Mooers}, \binits{A.O.}},
\bauthor{\bsnm{Pavoine}, \binits{S.}}:
\batitle{A guide to phylogenetic metrics for conservation, community evology
  and macroecology}.
\bjtitle{Biological Reviews}
\bvolume{92}(\bissue{2}),
\bfpage{698}--\blpage{715}
(\byear{2017})
\end{barticle}
\endbibitem

\bibitem[\protect\citeauthoryear{Hayward and
  Castley}{2018}]{HaywardCastley2018}
\begin{botherref}
\oauthor{\bsnm{Hayward}, \binits{M.W.}},
\oauthor{\bsnm{Castley}, \binits{J.G.}}:
Editorial: Triage in conservation.
Frontiers in Ecology and Evolution
\textbf{5}
(2018)
\doiurl{10.3389/fevo.2017.00168}
\end{botherref}
\endbibitem

\bibitem[\protect\citeauthoryear{Wiedenfeld
  et~al.}{2021}]{wiedenfeld2021conservation}
\begin{barticle}
\bauthor{\bsnm{Wiedenfeld}, \binits{D.A.}},
\bauthor{\bsnm{Alberts}, \binits{A.C.}},
\bauthor{\bsnm{Angulo}, \binits{A.}},
\bauthor{\bsnm{Bennett}, \binits{E.L.}},
\bauthor{\bsnm{Byers}, \binits{O.}},
\bauthor{\bsnm{Contreras-MacBeath}, \binits{T.}},
\bauthor{\bsnm{Drummond}, \binits{G.}},
\bauthor{\bsnm{Fonseca}, \binits{G.A.}},
\bauthor{\bsnm{Gascon}, \binits{C.}},
\bauthor{\bsnm{Harrison}, \binits{I.}}, \betal:
\batitle{Conservation resource allocation, small population resiliency, and the
  fallacy of conservation triage}.
\bjtitle{Conservation Biology}
\bvolume{35}(\bissue{5}),
\bfpage{1388}--\blpage{1395}
(\byear{2021})
\end{barticle}
\endbibitem

\bibitem[\protect\citeauthoryear{Bordewich and Semple}{2024}]{semple23}
\begin{barticle}
\bauthor{\bsnm{Bordewich}, \binits{M.}},
\bauthor{\bsnm{Semple}, \binits{C.}}:
\batitle{Quantifying the difference between phylogenetic diversity and
  diversity indices}.
\bjtitle{Journal of Mathematical Biology}
\bvolume{88}(\bissue{40}),
\bfpage{1}--\blpage{25}
(\byear{2024})
\end{barticle}
\endbibitem

\bibitem[\protect\citeauthoryear{Fischer et~al.}{2023}]{fischer23}
\begin{barticle}
\bauthor{\bsnm{Fischer}, \binits{M.}},
\bauthor{\bsnm{Francis}, \binits{A.}},
\bauthor{\bsnm{Wicke}, \binits{K.}}:
\batitle{Phylogenetic diversity rankings in the face of extinctions: {T}he
  robustness of the fair proportion index}.
\bjtitle{Systematic Biology}
\bvolume{72}(\bissue{3}),
\bfpage{606}--\blpage{615}
(\byear{2023})
\end{barticle}
\endbibitem

\bibitem[\protect\citeauthoryear{Manson and Steel}{2023}]{DIopt1}
\begin{botherref}
\oauthor{\bsnm{Manson}, \binits{K.}},
\oauthor{\bsnm{Steel}, \binits{M.}}:
Spaces of phylogenetic diversity indices: combinatorial and geometric
  properties.
Bulletin of Mathematical Biology
\textbf{85}
(2023)
\end{botherref}
\endbibitem

\bibitem[\protect\citeauthoryear{Manson}{2024}]{manson24}
\begin{barticle}
\bauthor{\bsnm{Manson}, \binits{K.}}:
\batitle{The robustness of phylogenetic diversity indices to extinctions}.
\bjtitle{Journal of Mathematical Biology}
\bvolume{89}(\bissue{1}),
\bfpage{5}
(\byear{2024})
\end{barticle}
\endbibitem

\bibitem[\protect\citeauthoryear{Moulton
  et~al.}{2024}]{moulton2024phylogenetic}
\begin{barticle}
\bauthor{\bsnm{Moulton}, \binits{V.}},
\bauthor{\bsnm{Spillner}, \binits{A.}},
\bauthor{\bsnm{Wicke}, \binits{K.}}:
\batitle{Phylogenetic diversity indices from an affine and projective
  viewpoint}.
\bjtitle{Bulletin of Mathematical Biology}
\bvolume{86}(\bissue{8}),
\bfpage{103}
(\byear{2024})
\end{barticle}
\endbibitem

\bibitem[\protect\citeauthoryear{}{2017}]{EDGEofExistence_2017}
\begin{botherref}
EDGE of Existence Programme
(2017).
\url{https://www.edgeofexistence.org/projects/}
\end{botherref}
\endbibitem

\bibitem[\protect\citeauthoryear{Faith}{1992}]{faith92}
\begin{barticle}
\bauthor{\bsnm{Faith}, \binits{D.P.}}:
\batitle{Conservation evaluation and phylogenetic diversity}.
\bjtitle{Biological Conservation}
\bvolume{61}(\bissue{1}),
\bfpage{1}--\blpage{10}
(\byear{1992})
\end{barticle}
\endbibitem

\bibitem[\protect\citeauthoryear{Ricotta}{2005}]{ricotta2005through}
\begin{barticle}
\bauthor{\bsnm{Ricotta}, \binits{C.}}:
\batitle{Through the jungle of biological diversity}.
\bjtitle{Acta biotheoretica}
\bvolume{53},
\bfpage{29}--\blpage{38}
(\byear{2005})
\end{barticle}
\endbibitem

\bibitem[\protect\citeauthoryear{Faith}{1994}]{faith1994biodiversity}
\begin{barticle}
\bauthor{\bsnm{Faith}, \binits{D.P.}}:
\batitle{Biodiversity and systematics: the use and misuse of divergence
  information in assessing taxonomic diversity}.
\bjtitle{Pacific Conservation Biology}
\bvolume{1}(\bissue{1}),
\bfpage{53}--\blpage{57}
(\byear{1994})
\end{barticle}
\endbibitem

\bibitem[\protect\citeauthoryear{Daly et~al.}{2018}]{daly2018ecological}
\begin{barticle}
\bauthor{\bsnm{Daly}, \binits{A.J.}},
\bauthor{\bsnm{Baetens}, \binits{J.M.}},
\bauthor{\bsnm{De~Baets}, \binits{B.}}:
\batitle{Ecological diversity: measuring the unmeasurable}.
\bjtitle{Mathematics}
\bvolume{6}(\bissue{7}),
\bfpage{119}
(\byear{2018})
\end{barticle}
\endbibitem

\bibitem[\protect\citeauthoryear{Redding
  et~al.}{2008}]{redding2008evolutionarily}
\begin{barticle}
\bauthor{\bsnm{Redding}, \binits{D.W.}},
\bauthor{\bsnm{Hartmann}, \binits{K.}},
\bauthor{\bsnm{Mimoto}, \binits{A.}},
\bauthor{\bsnm{Bokal}, \binits{D.}},
\bauthor{\bsnm{DeVos}, \binits{M.}},
\bauthor{\bsnm{Mooers}, \binits{A.{\O}.}}:
\batitle{Evolutionarily distinctive species often capture more phylogenetic
  diversity than expected}.
\bjtitle{Journal of theoretical biology}
\bvolume{251}(\bissue{4}),
\bfpage{606}--\blpage{615}
(\byear{2008})
\end{barticle}
\endbibitem

\bibitem[\protect\citeauthoryear{Steel}{2016}]{steel16}
\begin{bbook}
\bauthor{\bsnm{Steel}, \binits{M.}}:
\bbtitle{Phylogeny: Discrete and Random Processes in Evolution}.
\bpublisher{SIAM},
\blocation{Philadelphia, PA, USA}
(\byear{2016})
\end{bbook}
\endbibitem

\bibitem[\protect\citeauthoryear{Redding and Mooers}{2006}]{redding06}
\begin{barticle}
\bauthor{\bsnm{Redding}, \binits{D.W.}},
\bauthor{\bsnm{Mooers}, \binits{A.}}:
\batitle{Incorporating evoluationary measures into conservations
  prioritization}.
\bjtitle{Conservation Biology}
\bvolume{20}(\bissue{6}),
\bfpage{1670}--\blpage{1678}
(\byear{2006})
\end{barticle}
\endbibitem

\bibitem[\protect\citeauthoryear{Kouvelis and Gang}{2013}]{yu2013}
\begin{bbook}
\bauthor{\bsnm{Kouvelis}, \binits{P.}},
\bauthor{\bsnm{Gang}, \binits{Y.}}:
\bbtitle{Robust Discrete Optimization and Its Applications}
vol. \bseriesno{14}.
\bpublisher{Springer},
\blocation{Berlin, Germany}
(\byear{2013})
\end{bbook}
\endbibitem

\bibitem[\protect\citeauthoryear{Hartmann}{2013}]{hartmann13}
\begin{barticle}
\bauthor{\bsnm{Hartmann}, \binits{K.}}:
\batitle{The equivalence of two phylogenetic biodiversity measures: the shapley
  value and fair prop}.
\bjtitle{Journal of Mathematical Biology}
\bvolume{67}(\bissue{5}),
\bfpage{1163}--\blpage{1170}
(\byear{2013})
\end{barticle}
\endbibitem

\bibitem[\protect\citeauthoryear{Khalid}{2017}]{khalid17}
\begin{bbook}
\bauthor{\bsnm{Khalid}, \binits{S.}}:
\bbtitle{Introduction to Data Compression}.
\bpublisher{Morgan Kaufmann},
\blocation{Burlington, MA, USA}
(\byear{2017})
\end{bbook}
\endbibitem

\bibitem[\protect\citeauthoryear{Frohn and Manson}{2024}]{frohn24}
\begin{botherref}
\oauthor{\bsnm{Frohn}, \binits{M.}},
\oauthor{\bsnm{Manson}, \binits{K.}}:
A minimal compact description of the diversity index polytope.
arxiv.org/abs/2409.15641
(2024)
\end{botherref}
\endbibitem

\bibitem[\protect\citeauthoryear{Stadler}{2011}]{stadler11}
\begin{barticle}
\bauthor{\bsnm{Stadler}, \binits{T.}}:
\batitle{Simulating trees with a fixed number of extant species}.
\bjtitle{Systematic Biology}
\bvolume{60},
\bfpage{676}--\blpage{684}
(\byear{2011})
\end{barticle}
\endbibitem

\bibitem[\protect\citeauthoryear{Jetz et~al.}{2012}]{jetz12}
\begin{barticle}
\bauthor{\bsnm{Jetz}, \binits{W.}},
\bauthor{\bsnm{Thomas}, \binits{G.H.}},
\bauthor{\bsnm{Joy}, \binits{J.B.}},
\bauthor{\bsnm{Hartmann}, \binits{K.}},
\bauthor{\bsnm{Mooers}, \binits{A.O.}}:
\batitle{The global diversity of birds in space and time}.
\bjtitle{Nature}
\bvolume{491},
\bfpage{444}--\blpage{448}
(\byear{2012})
\end{barticle}
\endbibitem

\bibitem[\protect\citeauthoryear{G.~Barry~Baker and
  Wilkinson}{2002}]{Baker01032002}
\begin{barticle}
\bauthor{\bsnm{G.~Barry~Baker}, \binits{S.H.} \bsuffix{Rosemary~Gales}},
\bauthor{\bsnm{Wilkinson}, \binits{V.}}:
\batitle{Albatrosses and petrels in australia: a review of their conservation
  and management}.
\bjtitle{Emu - Austral Ornithology}
\bvolume{102}(\bissue{1}),
\bfpage{71}--\blpage{97}
(\byear{2002})
\doiurl{10.1071/MU01036}
\end{barticle}
\endbibitem

\bibitem[\protect\citeauthoryear{Phillips
  et~al.}{2016}]{phillips2016conservation}
\begin{barticle}
\bauthor{\bsnm{Phillips}, \binits{R.A.}},
\bauthor{\bsnm{Gales}, \binits{R.}},
\bauthor{\bsnm{Baker}, \binits{G.}},
\bauthor{\bsnm{Double}, \binits{M.}},
\bauthor{\bsnm{Favero}, \binits{M.}},
\bauthor{\bsnm{Quintana}, \binits{F.}},
\bauthor{\bsnm{Tasker}, \binits{M.L.}},
\bauthor{\bsnm{Weimerskirch}, \binits{H.}},
\bauthor{\bsnm{Uhart}, \binits{M.}},
\bauthor{\bsnm{Wolfaardt}, \binits{A.}}:
\batitle{The conservation status and priorities for albatrosses and large
  petrels}.
\bjtitle{Biological Conservation}
\bvolume{201},
\bfpage{169}--\blpage{183}
(\byear{2016})
\end{barticle}
\endbibitem

\bibitem[\protect\citeauthoryear{Wicke and Steel}{2020}]{wicke20}
\begin{barticle}
\bauthor{\bsnm{Wicke}, \binits{K.}},
\bauthor{\bsnm{Steel}, \binits{M.}}:
\batitle{Combinatorial properties of phylogenetic diversity indices}.
\bjtitle{Journal of Mathematical Biology}
\bvolume{80}(\bissue{3}),
\bfpage{687}--\blpage{715}
(\byear{2020})
\end{barticle}
\endbibitem

\end{thebibliography}
\end{document}